%% file: VandenBerk.tex
\documentclass[preprint]{aastex}

\newcommand{\eg}{e.g.~}
\newcommand{\etal}{et~al.~}
\newcommand{\civ}{C\,{\sc iv} }
\newcommand{\siv}{Si\,{\sc iv} }
\newcommand{\mgii}{Mg\,{\sc ii} }
\newcommand{\feii}{Fe\,{\sc ii} }
\newcommand{\caii}{Ca\,{\sc ii} }
\newcommand{\lya}{Ly\,$\alpha$ \ }

\begin{document}

\title{QSOs and Absorption Line Systems Surrounding the Hubble Deep Field}
\author{Daniel E. Vanden Berk\altaffilmark{1,2},
 Chris Stoughton\altaffilmark{2}}
\affil{Fermilab, MS 127, PO Box 500, Batavia, IL 60510}
\email{danvb@sdss.fnal.gov, stoughto@sdss.fnal.gov}
\author{Arlin P. S. Crotts}
\affil{Columbia Univ., Dept. of Astronomy, 550 W. 120th St., New York,
 NY, 10027}
\email{arlin@astro.columbia.edu}
\author{David Tytler, David Kirkman}
\affil{UC, San Diego, Ctr. Astroph. \& Space Sci., SERF Bldg., Rm. 428,
 LaJolla, CA 92093}
\email{tytler@ucsd.edu, david@cass151.ucsd.edu}
\altaffiltext{1}{Also: McDonald
 Observatory, University of Texas, RLM 15.308,Austin, TX, 78712}
\altaffiltext{2}{Work supported by the US Department of Energy under contract
DE-ACO2-76CH03000.}

\begin{abstract}
We have imaged a $45\arcmin \times 45\arcmin$ area centered on the Hubble Deep
Field \mbox{(HDF)} in $UBVRI$ passbands, down to respective limiting magnitudes of
approximately $21.5, 22.5, 22.2, 22.2, {\rm and \ } 21.2$.
The principal goals of the survey are to identify QSOs and to map structure
traced by luminous galaxies and QSO absorption line systems in a wide volume
containing the HDF.  The area surveyed is $400$ times that of the HDF, and
$40$ times that of the HDF Flanking Fields.  We have selected QSO candidates
from color space, and identified 4 QSOs and 2 narrow emission-line
galaxies (NELGs) which have not previously been discovered, bringing the
total number of known QSOs in the area to $19$.  The bright $z=1.305$ QSO
only 12\arcmin \ away from the HDF raises the northern HDF to nearly the
same status as the HDF-S, which was selected to be proximate to a bright QSO.
About half of the QSO candidates remain for spectroscopic verification.

Absorption line spectroscopy has been obtained for $3$ bright QSOs in the
field, using the Keck 10m, ARC 3.5m, and MDM 2.4m telescopes.  Five
heavy-element absorption line systems have been identified, 4 of which
overlap the well-explored redshift range covered by deep galaxy redshift
surveys towards the HDF.  The two absorbers at $z=0.5565$ and $z=0.5621$
occur at the same redshift as the second most populated redshift peak in
the galaxy distribution, but each is more than $7h^{-1}$Mpc (comoving,
$\Omega_{m}=1$, $\Omega_{\Lambda}=0$) away from the HDF line of sight in the
transverse dimension. This supports more indirect evidence that the galaxy
redshift peaks are contained within large sheet-like structures which
traverse the HDF, and may be precursors to large-scale ``pancake''
structures seen in the present-day galaxy distribution.
\end{abstract}


\section{Introduction \label{intro}}
Deep galaxy redshift samples are permitting a new and often surprising view
of the Universe at much younger epochs, and into which  the role of gas, both
hydrogen and processed, via QSO absorption line systems can be incorporated.
Only recently, and with the help of the 10-m Keck telescopes, have deep galaxy
redshift surveys been able to measure properties of galaxies at some of the
redshifts ($2.5 \la z \la 4.5$) which have been easily accessible to absorption
line studies for over three decades.  Combining the study of QSO absorbers
and galaxy surveys has the potential to greatly enhance our understanding
of the formation and evolution galaxies as well as the large-scale structures
which typically contain them.
For example, even if galaxies and absorbers are closely related, biasing,
which  plays an important role in deciphering structure formation, is expected
to be different for for galaxies, QSOs, absorbers,
and the various classes of each \citep[\eg][]{demia1999,cen1998,fang1998,
quashnock1998,bi1996}.

A generic result of the deep galaxy pencil-beam surveys is that half or more of
the galaxies measured tend to lie in very narrow redshift ``spikes'' which
are present to redshifts of at least $z=1$ \citep{cohen1996a,cohen1996b} and
are often found at much higher redshifts ($z\approx 3$) in the ``dropout''
surveys \citep{steidel1998,adelberger1998}.  The number density, redshift
spacing, density
enhancements, velocity dispersions, and morphological mixtures, all support
the hypothesis that these structures in redshift space are parts of the
precursors to present-day galaxy superclusters and walls \citep{cohen1996a,
cohen1996b}.  This evidence is mostly circumstantial so far, since
the deep pencil-beam surveys cover only very small (typically 50 sq.\
arcmin.\ or less) disjoint areas
of the sky.  Additional but shallower redshift surveys have been carried out in
narrow fields adjacent to at least one deep pencil beam survey, which have
supported the the idea that the redshift structures are coherent in the
transverse spatial dimension on scales up to at least a degree, and for
redshifts up to at least $z\approx 0.4$ \citep{cohen1999}.
Extending this type of survey to deeper redshifts is difficult not only
due to the faintness of the galaxies, but in the redshift range
$1.2 \la z \la 2$ there is a lack of redshifted galaxy spectral features
available at optical wavelengths.
It is a highly desirable but currently difficult goal of future redshift
surveys to cover both larger areas and a more complete redshift range. 
QSO absorption line systems offer a means of efficiently extending these
studies to wider volumes and higher redshift, which is the aim of the
program described here.

The approach is to search for intervening absorption line systems in the
spectra of QSOs at small angular separations. The selection function for
heavy-element QSO absorption line systems,
identified mainly by \civ $\lambda$1550{\AA} and  \mgii $\lambda$2799{\AA}
doublet transitions, is luminosity independent, and limited at high redshift
only by the emission redshift of the backlighting QSOs.  In optical spectra
\mgii lines can be detected from redshifts of $z \approx 0.15 - 2.0$
and \civ lines from $z\approx 1$ to over 4.  Absorption surveys
towards groups of QSO sightlines have been successfully used to trace
structure in three dimensions at high redshift \citep{crotts1985,crotts1989,
jakobsen1992,foltz1993,elowitz1995,dinshaw1996,williger1996,vandenberk1999,
impey1999}.
A few QSOs have also been observed directly within the areas covered
by the galaxy surveys, and their spectra have revealed absorption line
systems that very often lie within the redshift peaks defined by the galaxies
\citep{steidel1998}.  These studies have demonstrated the utility
of absorption line systems in probing large-scale structure both in radial and
angular dimensions, and of using large-scale structure studies to decipher the
relationship between galaxies and absorbing gas.

QSOs bright enough to use for 3-dimensional absorption line studies generally
have a high enough angular density so that suitable groups can be found in 
virtually any field of sufficiently high galactic latitude.  For example, most
UVX QSO surveys reveal a density of about 30 QSOs per sq.\ degree to a limiting 
magnitude of $B\le 21$ \citep*[\eg][]{zhan1989}, which is a practical
limit for absorption line surveys with 4-m class telescopes. To take full
advantage of this technique, one should select fields
in which deep galaxy redshift surveys have also taken place.  The galaxy and
absorber surveys are then complementary: the galaxies provide the redshift
locations and velocity dispersions of structures, while the absorbers can
be used to quickly and efficiently widen the survey to larger areas and
additional redshift ranges, and probe the otherwise invisible structure of
the gas.

The Hubble Deep Field \citep[HDF;][]{williams1996} is the site of one of the
most complete and comprehensive sets of deep redshift surveys, with over
$300$ measured redshifts in an area of only $\approx 50$
sq.\ arcmin.\ \citep{cohen1996a,steidel1996,lowenthal1997,guzman1997,
phillips1997,hogg1998}.  The measured redshifts lie in the range $z \la 1.3$
and $z \ga 2.0$, with a gap between $1.3$ and $2.0$ due to restrictions of
optical spectroscopy.  We have chosen the area surrounding the Hubble Deep
Field for our initial QSO/absorber study because of the large and continuing
amount of research devoted to this sightline, and because it is easily
accessible not only by northern-hemisphere telescopes, but also to the Hubble
Space Telescope (HST) which can be used for follow-up observations of the
low-redshift \lya
systems.  Indeed, this latter approach is the primary justification for the 
construction of the Hubble Deep Field South, and a survey similar
to ours for additional QSOs in that direction of the sky is currently taking
place \citep{teplitz1998}.  \citet[hereafter LPIF]{lpif} recently carried
out a QSO survey in the one square degree surrounding the HDF, and found 30
QSOs brighter than $B=21$.  While the LPIF survey and ours have similar goals
and survey depths, ours uses 5-band photometry (LPIF used only $U, B,
{\rm and \ }
R$ bands) to search for high-redshift QSOs, and we have started QSO absorption
line follow-up spectroscopy.  Comparisons of the two surveys will be made when
appropriate.  In this paper we present our initial results on the QSO
survey towards the HDF (there are no reasonably bright QSOs inside the HDF
itself), and our preliminary absorption line study of 3 of
the QSO lines-of-sight.  The imaging observations and photometry, QSO
candidate selection and verification, and QSO absorption spectroscopy, are
presented in \S\,\ref{imandph}, \S\,\ref{candsec}, and \S\,\ref{qsoals}
respectively.  We discuss the distributions of the QSOs and absorbers
relative to the galaxy redshift sample in \S\,\ref{discsec}.  A summary is given
in \S\,\ref{summary}.

\section{$U,B,V,R,I$ Imaging and Photometry \label{imandph}}

\subsection{Observations and Image Reduction \label{imaging}}
Images centered on the Hubble Deep Field were taken at McDonald Observatory
using the Prime Focus Camera (PFC) mounted on the 0.76 m telescope.  The
PFC is a dedicated prime focus (f/3.0) corrector with a $2048 \times 2048$  
Loral Fairchild CCD, which covers an area of $46.25 \times 46.25$ sq.\
arcminutes (a plate scale of $1.355$ arcsec$/$pixel).
One CCD field, centered on the HDF, was imaged many times in each of the
five filters of the Bessel $UBVRI$ system.  Small ($\sim 50$ pixel) offsets
were made between each of the exposures to facilitate the removal of CCD chip
defects and cosmic ray events.

The observations were made on several nights in late February and early
March, 1998.  The seeing was exceptionally good on two of the nights, yielding
point spread functions with typical FWHM less than 2 pixels.  These were also
the only photometric nights, such that the standard star observations were
useful for absolute photometric calibrations.  About half of the images in
each band were taken in these conditions.
The seeing was substantially worse during the other nights, and it turned
out that the co-addition of frames taken on those nights did not improve
the image depths enough to justify the loss of morphological information and
close-source separation.
The primary goal of the QSO search is to identify QSOs bright enough for
absorption line system spectroscopy follow-up, so the marginal improvement
in magnitude limits is not ultimately important.

The raw images were reduced using a package of IRAF\footnote{IRAF is 
written and supported by the IRAF programming group at the National Optical
Astronomy Observatories (NOAO) in Tucson, Arizona.} scripts written by
Inger J{\o}rgensen specifically to reduce McDonald PFC imaging data.
Individual science frames were corrected for bias level, differential
shutter open time, flat fields, and illumination gradients.  Science frames
in each band were co-added taking into account seeing, bad pixels, background
level, and noise in each frame.
The total exposure times and FWHM for the final coadded science images is
given in Table~\ref{imagelog}.

\subsection{Photometry \label{photom}}
Standard star observations were taken each night the sky appeared to
be photometric.  The photometric stability was acceptable on only two of the
nights.  Typically $4-5$ Johnson-Kron-Cousins $UBVRI$ standard
star fields from the list of \citet{landolt1992} were observed each night at
high and low airmasses.  The standard star frames were reduced in the
same way as the science frames (but with no coaddition).  Aperture
photometry was performed on all of the standard stars in the frames,
and an aperture correction was determined for each filter.

Zero point offsets ($u_0, b_0, v_0, r_0, i_0$), extinction coefficients
($u_1, b_1, v_1, r_1, i_1$), and color corrections ($u_2, b_2, v_2, r_2, i_2$)
were determined for the two photometric nights, by interactively fitting the
parameters of sets of equations like those below.  The equations relate the 
aperture-corrected instrumental magnitudes ($u,b,v,r,i$) to the standard
Johnson-Kron-Cousins magnitudes ($U,B,V,R,I$), the airmass of the observations
($X_u,X_b,X_v,X_r,X_i$), and a color term:
\begin{eqnarray}
  u & = & U + u_0 + u_1 \times X_u + u_2 \times (U - B) \\
  b & = & B + b_0 + b_1 \times X_b + b_2 \times (B - V) \\
  v & = & V + v_0 + v_1 \times X_v + v_2 \times (V - R) \\
  r & = & R + r_0 + r_1 \times X_r + r_2 \times (V - R) \\
  i & = & I + i_0 + i_1 \times X_i + i_2 \times (R - I)
\end{eqnarray}

Systems of equations with many different color terms were fit, since not
all science objects were detected in every co-added image.
The equations were transformed to yield functions for each of the 
standard magnitudes.  The coefficients varied slightly between the
two nights, so separate transformations were applied to the data
taken on each of the nights.  No correction was made for Galactic reddening,
since the direction towards the HDF has a very small reddening factor
\citep{williams1996}.

Objects in the co-added science frames were detected, and their fluxes
measured, using the program SExtractor \citep{bertin1996}.
SExtractor determines the background, detects objects, deblends multiple
sources extracted as single objects, measures magnitudes, and 
discriminates between point-like and extended objects.  The program
parameters were adjusted so that all objects clearly identified as
separate objects by eye were detected and deblended with SExtractor.
The SExtractor ``best estimate'' \citep[using either adaptive aperture or
corrected isophotal photometry;][]{bertin1996} of the total flux for each
object, was used for magnitude calculations.  SExtractor was also run on the
standard star frames in order to compare the aperture-corrected and
SExtractor magnitude estimates.  There was a small ($<3\%$) offset
between the two estimates for the standard stars, which did not appear
to be magnitude dependent; this offset was applied to the SExtractor
magnitudes of the science objects.

An effective airmass for each co-added image was determined by fitting a
line to the individual science frame airmasses vs.\ the instrumental
magnitude differences between the co-added frame and the individual frames.
The effective airmass (the $y$-intercept of the line fit) for each
co-added image was used in the photometric transformation equations.

Science objects were matched on all co-added frames in which they were detected.
The final standard magnitude determinations were made for each science object,
using the transformation equations, including the color term if possible.
If an object was detected in only one frame (for example, faint objects
in the $R$ band image, or objects close to a non-overlapping frame edge),
no color term was applied.   

\begin{figure}
\epsscale{0.8}
\plotone{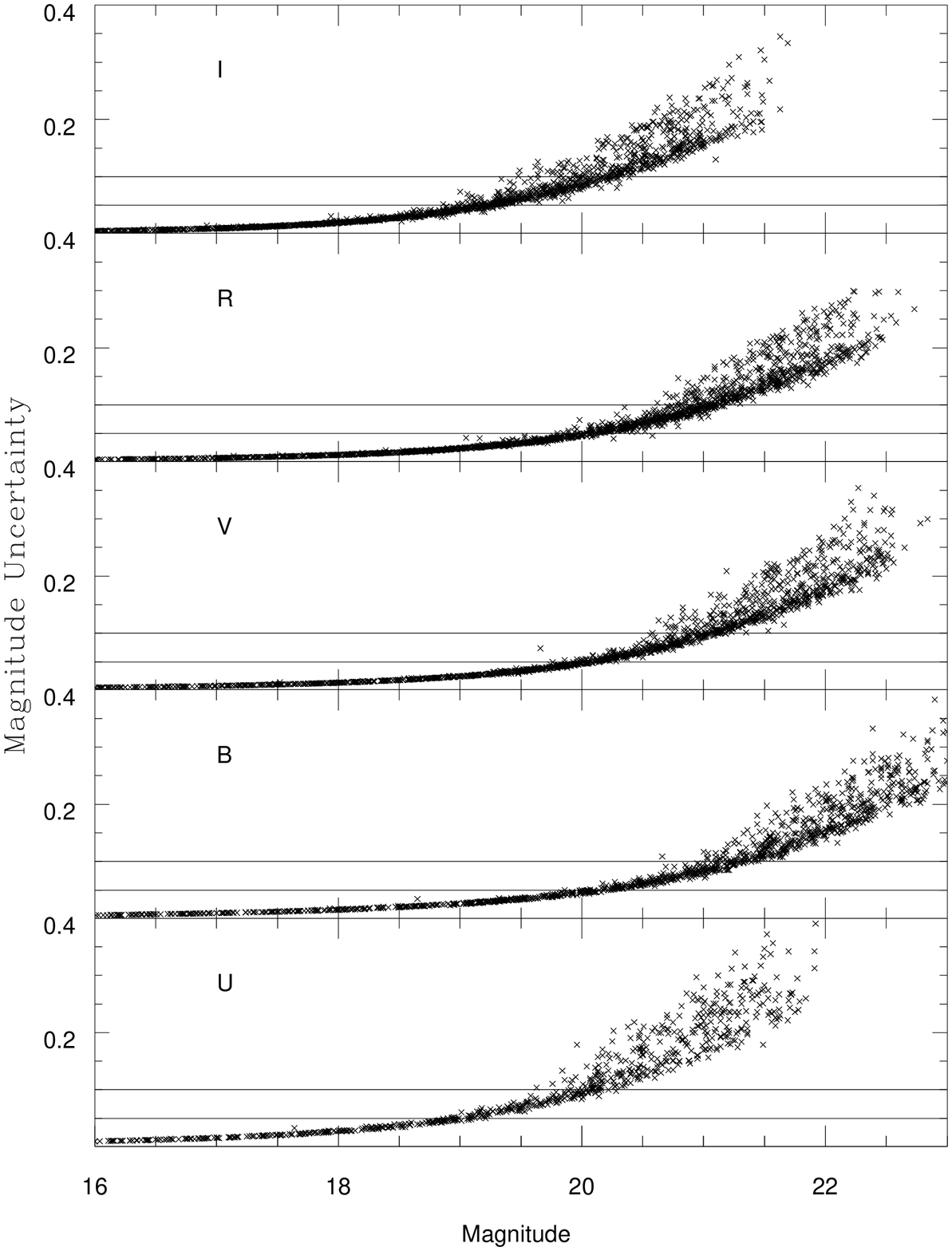}
\caption[fig01.eps]{Magnitude uncertainties for all objects
  with stellarity index greater then 0.6, separated by passband. \label{magerr}}
\end{figure}

Uncertainties in the magnitude estimates are given by SExtractor, based
upon the flux and extent of the object, and the background variance.
These estimates agreed well with uncertainties based upon variations among
individual science frames, except for saturated objects.
Stars become saturated in our images at approximately $U=12.3, \ B=14.5,
 \ V=13.8, \ R=14.1, \ {\rm and} \ I = 14.1$.
We have used the SExtractor estimates for the magnitude uncertainties
except for objects brighter than the saturation level,
which we simply flagged as ``saturated''.
The magnitude uncertainties, shown in Fig.\,\ref{magerr}, reach the 10\%
level at $U=20.1, \ B=21.3, \ V=21.1, \ R=21.1, \ {\rm and} \ I=20.2$. 

Aside from the statistical uncertainties, there may be systematic
uncertainties in the magnitude estimates due to a variety of possible
causes.  For our purposes, systematic uncertainties are not worrisome
as long as the stellar locus is well defined by the measured colors, and
outliers can be easily identified.  However, accurate magnitudes in 
an absolute sense are necessary for other uses of the data, and for
comparison with other studies.  To check this, we have compared our
$UBR$ magnitudes with those of LPIF (who did not take observations in
the $V$ or $I$ bands), for objects common to the two
lists.  The $B$ and $R$ band measurements in each set are not significantly
different, however, our measured $U$ magnitudes are brighter on average by
0.13 mag, which is about 4 standard errors of the mean away from
no difference in the $U$ measurements.
The vast majority of the objects used for comparison are
fainter than $U=20.1$, the 10\% uncertainty level of our $U$ band data,
which may account for the larger difference. It is also possible that since
all of our $U$ band observations were done on a single night, the difference
can be attributed to uncorrected sky variations.  In any case, the color-color
diagrams (\S\,\ref{candsec}) appear to be in good agreement with those of other
studies, and we have not applied additional photometric corrections.

\subsection{Astrometry \label{astrom}}
Astrometry was performed by comparing the pixel coordinates (determined using
SExtractor) of stars in
our field to the J2000 equatorial coordinates given in the HST Guide Star
Catalog\footnote{The Guide Star Catalog was produced at the Space Telescope
Science Institute under U.S. Government grant. These data are based on
photographic data obtained using the Oschin Schmidt Telescope on Palomar
Mountain and the UK Schmidt Telescope.}.  About 90 GSC stars appear in the
co-added images.  Tasks in the IRAF imcoords package were used to fit a
2-dimensional polynomial function to the pixel/equatorial coordinates.  The
r.m.s.\ of the residuals was less than $0.5$ arcseconds in both the $x$ and $y$
pixel dimensions for each co-added image.

\subsection{Star-Galaxy Separation \label{stargal}}
Discrimination between point-like or extended objects is a serious issue
for our dataset, given the fairly large pixel size ($1.355\arcsec/$pix).
The method we used was based upon the ``stellarity index'' produced by the
SExtractor neural-network classifier \citep{bertin1996}.
Each object detected in a co-added image was assigned a number between 0
and 1, which represents the confidence that an object is stellar.
The stellarity index match for objects detected in multiple bands was quite
good. For example, for objects detected in both the $U$ and $R$ band, the
indices matched to within 0.1 for more than 75\% of the objects, although the
correlation degrades with fainter magnitudes.  For multiply detected objects,
the indices were weighted by the inverse of the squared magnitude uncertainties,
then averaged over all the detections.
We used an index $\geq 0.6$, because this included a large number of
objects without noticeably increasing the width of the stellar locus in
the color-color diagrams.  In addition, all but one of the confirmed stars
and QSOs from LPIF are selected with this cut.

\subsection{The Object Catalog \label{catalog}}
The final imaging catalog contains 10647 objects detected in at least one 
band, 1516 detected in all 5 bands, and 1836, 4033, 5681, 8736, and 7841
objects detected in the $U, \ B, \ V, \ R, \ {\rm and} \ I$ bands respectively.
There are 2147 objects classified as stellar, or about 20\% of the total
number of objects.  Objects selected as QSO candidates will be presented
in the next section, but the full catalog is likely to be useful for other
studies, particularly because the survey area contains and surrounds the
HDF.  The full object catalog containing coordinates, magnitudes,
uncertainties, and stellarity indices for all of the detected objects, may
be obtained by contacting the authors.

\section{QSO Candidate Selection and Verification \label{candsec}}

\subsection{Selection of QSO Candidates}
QSO candidates were selected based upon their locations in color space.
In order to maintain a reasonably high efficiency of QSO selection,
the candidates should clearly be located well outside of the stellar
locus, and in regions of color space which QSOs are known to occupy at a 
relatively high density.
The simplest selection criterion is to make one or two-dimensional cuts in
two-color spaces.  For example, QSOs with redshifts up to $\sim 2.2$
can be found with good efficiency by selecting objects with $U - B < -0.3$.
At various redshifts, strong features in QSO spectra, such as the
onset of the Ly\,$\alpha$ forest, move into and out of different photometric
passbands.  Thus QSO colors can be strongly dependent on redshift, and
various color space cuts are most efficacious over select ranges of redshift.
Automated but more complex outlier selection techniques have been tried in
other studies \citep[\eg][]{newberg1997,warren1991}, which
are appropriate for large surveys for which it is impractical to check every
candidate.  Our sample is small enough to check
each outlier individually, and our goal is not to identify a complete
sample of QSOs, so color space cuts, based in part on results from past
studies, are adequate for our purposes.

\begin{figure}
\epsscale{0.8}
\plotone{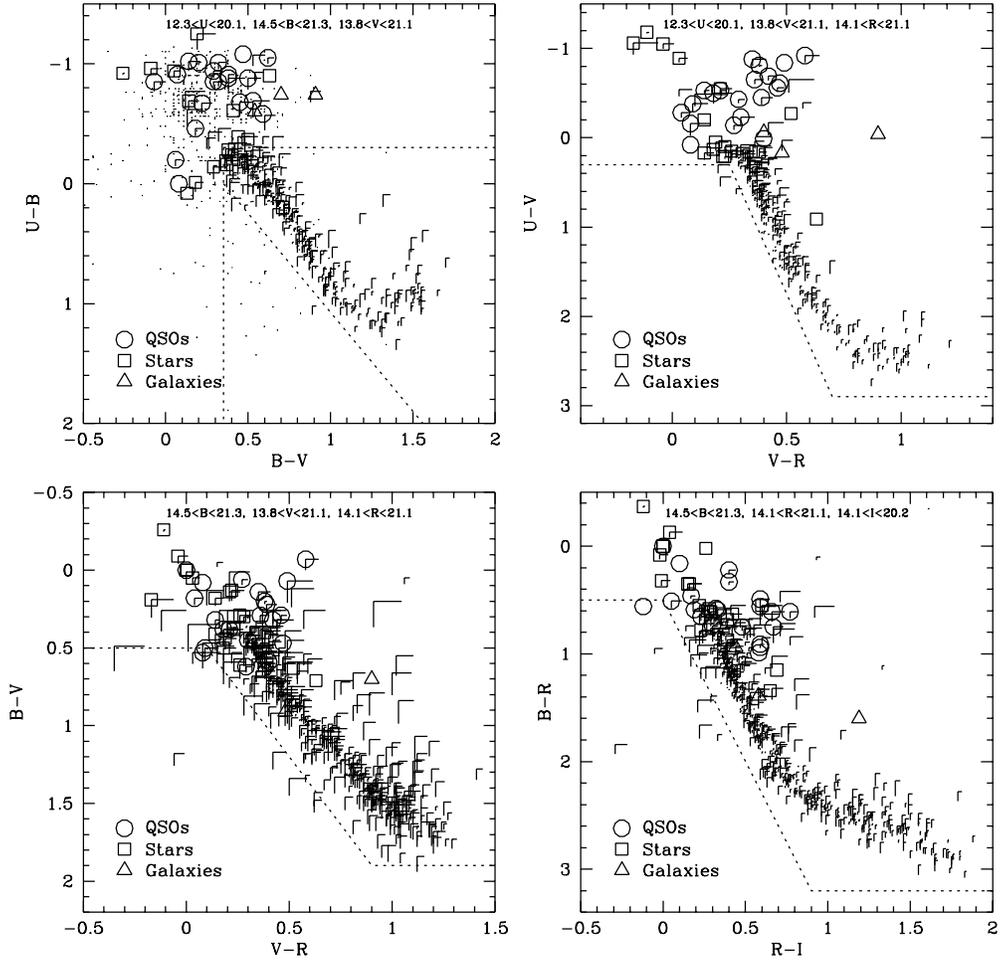}
\caption[fig02.eps]{Color-color plots for point-like objects
  which are brighter than the 10\% uncertainty level in all 3 passbands.  The
  $1\sigma$ uncertainties in each color are shown in only two directions for
  clarity. Dashed lines show the boundaries used for selecting QSO candidates.
  The color-space locations of spectroscopically confirmed QSOs, NELGs, and
  stars are shown by circles, triangles, and squares respectively. \label{cm}}
\end{figure}

To select the search regions, two-color diagrams were plotted for all
unsaturated objects classified as stellar, with magnitudes brighter than
the 10\% uncertainty level (Fig.\,\ref{cm}).  For the ($U-B)/(B-V$) diagram,
we also plotted all of the available colors of QSOs in the catalog of
\citet{veron1996}.  Based on the location of the stellar
locus and the known QSOs, cuts in ($U-B)/(B-V$) color space were made to
select candidates from our imaging catalog.  An historically successful
method for selecting QSOs up to $z\lesssim2.2$ is to select objects with
$U-B \leq -0.3$ -- the so-called ``UV-excess'' method.  This also appeared
to be a good cutoff for our dataset, as seen in Fig.\,\ref{cm}.  In addition,
a large fraction of the QSOs plotted from the V{\' e}ron-Cetty \&
V{\' e}ron catalog could have been selected
with a $B-V \leq 0.35$ cut, which includes only a small part of the stellar
locus (mostly A-stars).  These combined cuts in the ($U-B)/(B-V$) plane
are the ``UVX'' selection method.  UVX candidates with $B-V>0.6$ are often
identified as compact narrow emission line galaxies (NELGs) instead of QSOs.
Few of our candidates exceed this limit, so we have not used it as a
selection criterion.  UVX candidates were divided into ``bright'' and
``faint'' sets, depending on whether the $U,B, \ {\rm and} \ V$ magnitudes
were brighter or fainter than the 10\% uncertainty levels.  It is assumed
that the ``bright'' candidates have a higher fraction of true QSOs, due
to their better photometric accuracy.  The UVX candidates that have not
been spectroscopically confirmed are listed in Table~\ref{qsouvx}.

Other cuts in color space, mainly aimed at locating higher-redshift QSOs,
have been explored in other studies \citep[\eg][]{irwin1991,hall1996}.
The number density of QSOs with $z>3$ to the limits of
our survey is only about $\sim 5$ per square degree \citep[\eg][]{hall1996},
but even one high-redshift QSO in the direction
of the HDF would be valuable for absorption system studies.  Hall et~al.,
using a filter set similar to ours, successfully used cuts in color
space to identify QSOs up to $z=4.33$.  We have adopted several of their
selection criteria, with slight modifications, in order to search for
higher-redshift candidates in our catalog.  These cuts are based on the
($U-V)/(V-R$), ($B-V)/(V-R$), and
($B-R)/(R-I$) two-color diagrams, and usually include objects which are red
in the first color and blue in the second.  In addition, we have added a
cut in the ($U-B)/(B-V$) plane for objects red in $U-V$ which are also blue
in $B-V$.  The color selection criteria are shown in Fig.\,\ref{cm} and listed
in Table~\ref{qsocut}.  The cuts are set at reasonable values designed to
run close to the stellar locus without introducing a large fraction of stars.
We call these criteria collectively the ``high-$z$'' selection methods.
As with the UVX selection, we have divided the high-$z$ sample into 
``bright'' and ``faint'' sets, according to the 10\% magnitude uncertainty
levels.  Because larger photometric errors in the faint set can cause 
a large number of non-QSOs to be selected as candidates, we kept only those
faint high-$z$ candidates which passed as candidates using more than one
two-color cut.  As expected, the high-$z$ selection did not produce as many
QSO candidates as the UVX selection.  The high-$z$ candidates which have not
been spectroscopically identified are listed in Table~\ref{qsohiz}.

\subsection{Spectroscopic Candidate Verification \label{verify}}
The initial candidate verification was done as a poor-weather contingency
program in April 1998 using the McDonald Observatory 2.7m telescope and Large
Cass Spectrometer (LCS).  All $7$ of the QSO candidates we observed in this run
have also been observed by LPIF, and the $4$ QSOs and 1 NELG
we confirmed in this run were also identified by them.  A fifth candidate
(J123800+6213) also turned out to be a QSO, but the S/N level in our spectrum
was too low to identify it as such.
For a second verification run in March 1999, we had the advantage of the
published QSO list of LPIF, and so were able to avoid candidates which
had already been observed.  Our candidate list contained many UV-excess
objects and high-$z$ candidates not in any of the lists of LPIF.
We used the McDonald 2.7m and IGI spectrograph to observe $11$ candidates,
which yielded 4 QSOs and 2 NELGs.

\begin{figure}
\epsscale{1.0}
\plottwo{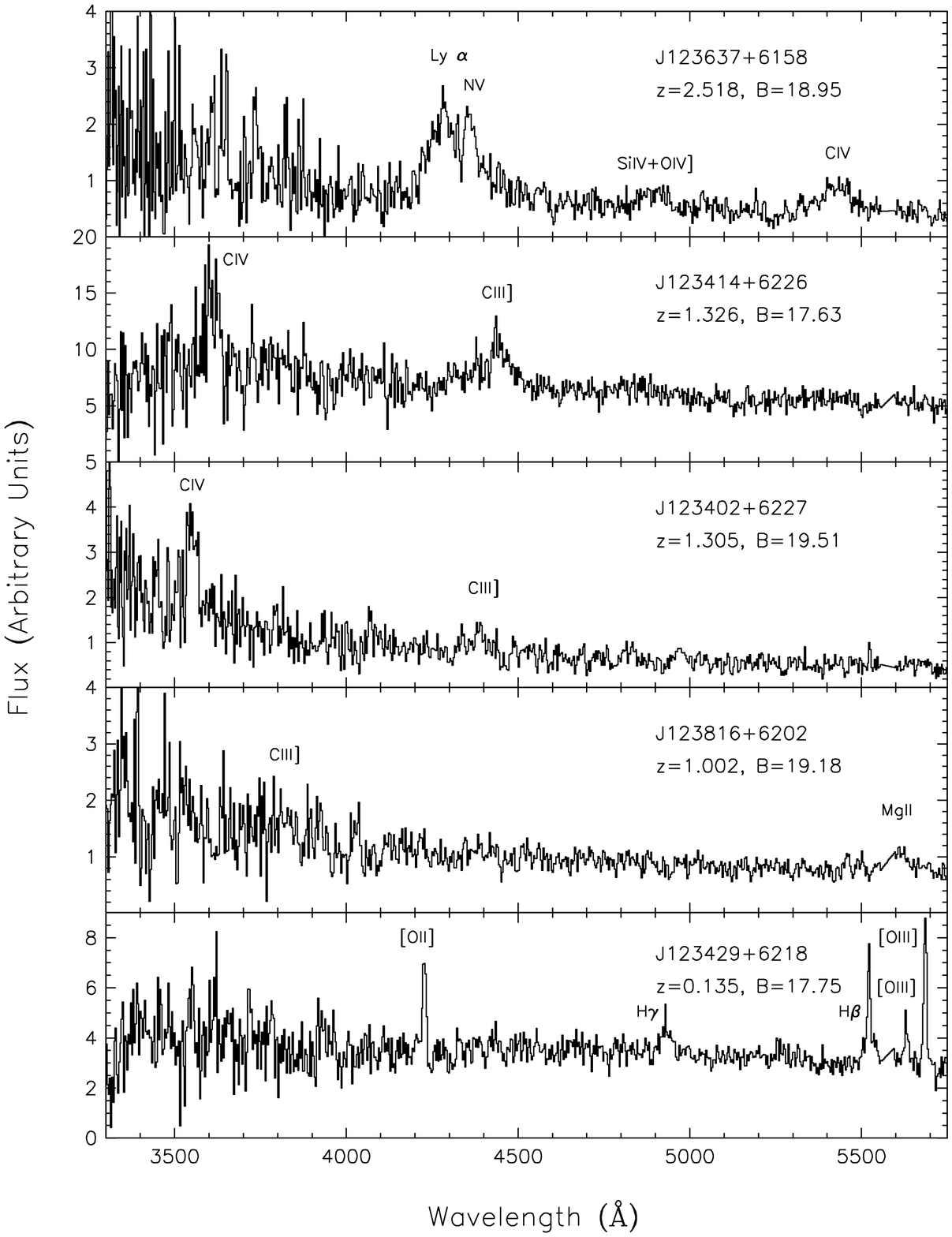}{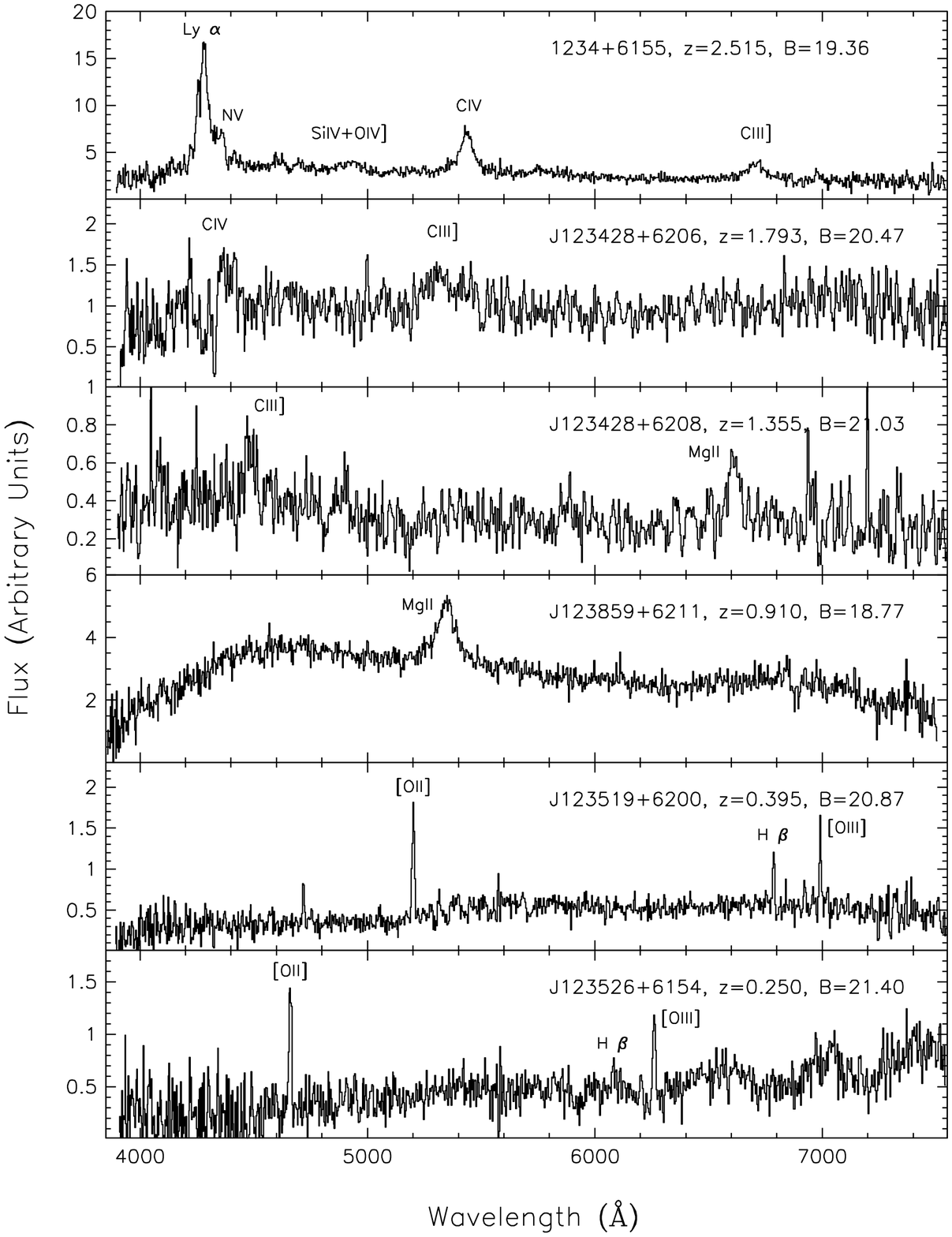}
\caption[fig03a.eps,fig03b.eps]{Spectra of QSOs and
  NELGs confirmed in this program. Spectra from the April 1998 run are on the
  left, and from the March 1999 run are on the right. The spectra of
  J123428+6206 and J123428+6208 have been smoothed by 2 pixels to improve the
  $S/N$ ratios. \label{qsospec}}
\end{figure}

All of the spectra were reduced using standard techniques and tasks in IRAF.
Bias level and flat field corrections were applied, then the spectra were
optimally extracted \citep{horne1986}, wavelength calibrated, and co-added.
A sensitivity correction was applied to give an indication of the relative
spectral shapes, but the spectra were not flux calibrated. 
The final reduced, co-added, calibrated QSO and NELG spectra are shown in
Fig.\,\ref{qsospec}.

\begin{figure}
\epsscale{0.8}
\plotone{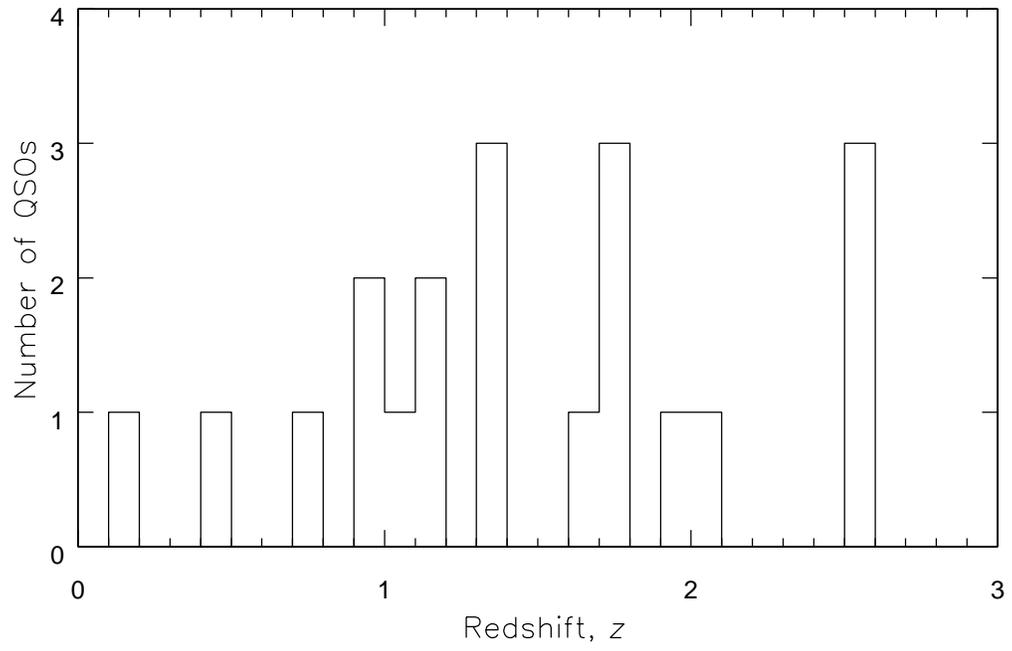}
\caption[fig04.eps]{The redshift distribution of the 19 QSOs
  and 1 AGN identified within our survey area. \label{zdist}}
\end{figure}

\begin{figure}
\epsscale{0.8}
\plotone{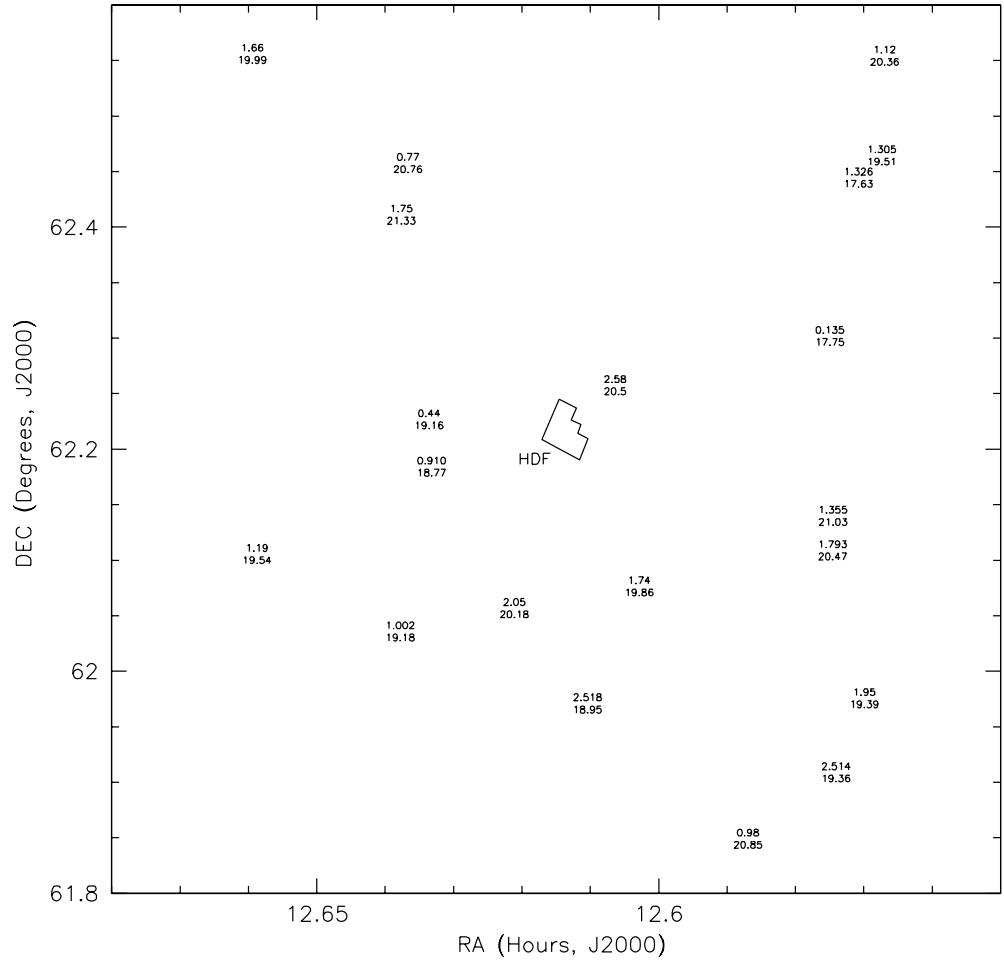}
\caption[fig05.eps]{Celestial coordinate positions (J2000) of
  QSOs identified within our survey area.  The QSOs are labeled by their
  redshifts and $B$ magnitudes.  The location of the HDF WFPCII area is
  indicated. \label{coords}}
\end{figure}

In total, we observed $18$ QSO candidates, of which $9$ are QSOs,
$1$ is an AGN, $2$ are narrow-emission-line galaxies,
$4$ are stars, and $2$ remain unidentified.  An additional 19 objects in our
candidate list were observed by LPIF, $9$ of which are QSOs, and $10$
of which are are stars.  One QSO from the LPIF list, J123622+6215, was not
selected as a candidate by us, since it has a stellarity index of only
$0.37$ caused by blending with a fainter object.  The QSOs and
NELGs confirmed in our program and
those confirmed by LPIF in our survey area, are listed in Table~\ref{qsoid}.
The identified stars are listed in Table~\ref{stars}.  There are 16 more
bright UV-excess candidates in our list which we have not observed
spectroscopically, and which do not appear in the candidate list of LPIF.  The
unconfirmed UVX QSO candidates are listed in Table~\ref{qsouvx}, and the
unconfirmed high-$z$ candidates are listed in Table~\ref{qsohiz}.
The redshift distribution of the identified QSOs is shown in
Fig.\,\ref{zdist}, and the coordinate positions are shown in
Fig.\,\ref{coords}.

While our goal is not a complete survey of QSOs in the area, it is 
useful to compare the density of QSOs near the HDF to other QSO
surveys.  Including the QSOs found by us and by LPIF in the
$\sim0.56$ sq.\ deg.\
area surrounding the HDF, there is a total of $17$, or roughly $30$ per
square degree down to a limiting magnitude of $B=21$.  This would be in
good agreement with the densities found in several other faint UVX surveys
\citep[\eg][]{koo1988,zhan1989}, which find roughly $30$ per sq.\ deg.,
except that we have a remaining 32 unidentified candidates with $B\leq21$.
Applying our UVX success rate (just over 50\%, including the results of
LPIF) to the remaining 32 UVX candidates with
$B\leq21$, we expect about another $16$ QSOs in our candidate list.
Many of the remaining UVX candidates lie close to the $U-B$ and $B-V$
selection limits, and some lie in the region more heavily populated by
NELGs, so it is doubtful that the efficiency for the remaining candidates
will be as high as $50\%$.  Assuming an efficiency only half this ($25\%$)
we would reasonably expect about another $8$ QSOs, making the UVX QSO
density about $45$ per sq.\ deg.\ to $B\leq21$. While this is significantly
higher than most previous studies, the density is in good agreement with
\citet{hall1996} who also noted that the density they found (in separate
survey areas) was surprisingly high.  The discrepancy may be due to
differences in CCD vs.\ photographic detection techniques, some other
selection difference, or real differences in the QSO
number density, but the issue is unresolved.  In any case, we conclude that
our candidate selection is both relatively complete and efficient, and
more than adequate for our purposes.

\section{QSO Absorption Line System Spectroscopy \label{qsoals}}

The principal goal of the QSO survey is to provide targets for
higher-resolution follow-up spectroscopy in order to locate QSO absorption
line systems near the HDF.  After our first verification run, we identified
4 bright QSOs.
Spectra suitable for absorption system searches were obtained for 3 of them:
J123414+6226 ($z=1.326$), J123402+6227 ($z=1.305$), and J123637+6158
($z=2.518$).  One of these (J123414+6226) was bright enough to observe at
very high resolution using the Keck HIRES spectrograph \citep{vogt1994}.
QSOs J123414+6226 and J123402+6227 are separated by only 112 arcsec.
The observing logs for the higher resolution spectra are summarized in
Table~\ref{hireslog}.

\begin{figure}
\epsscale{0.8}
\plotone{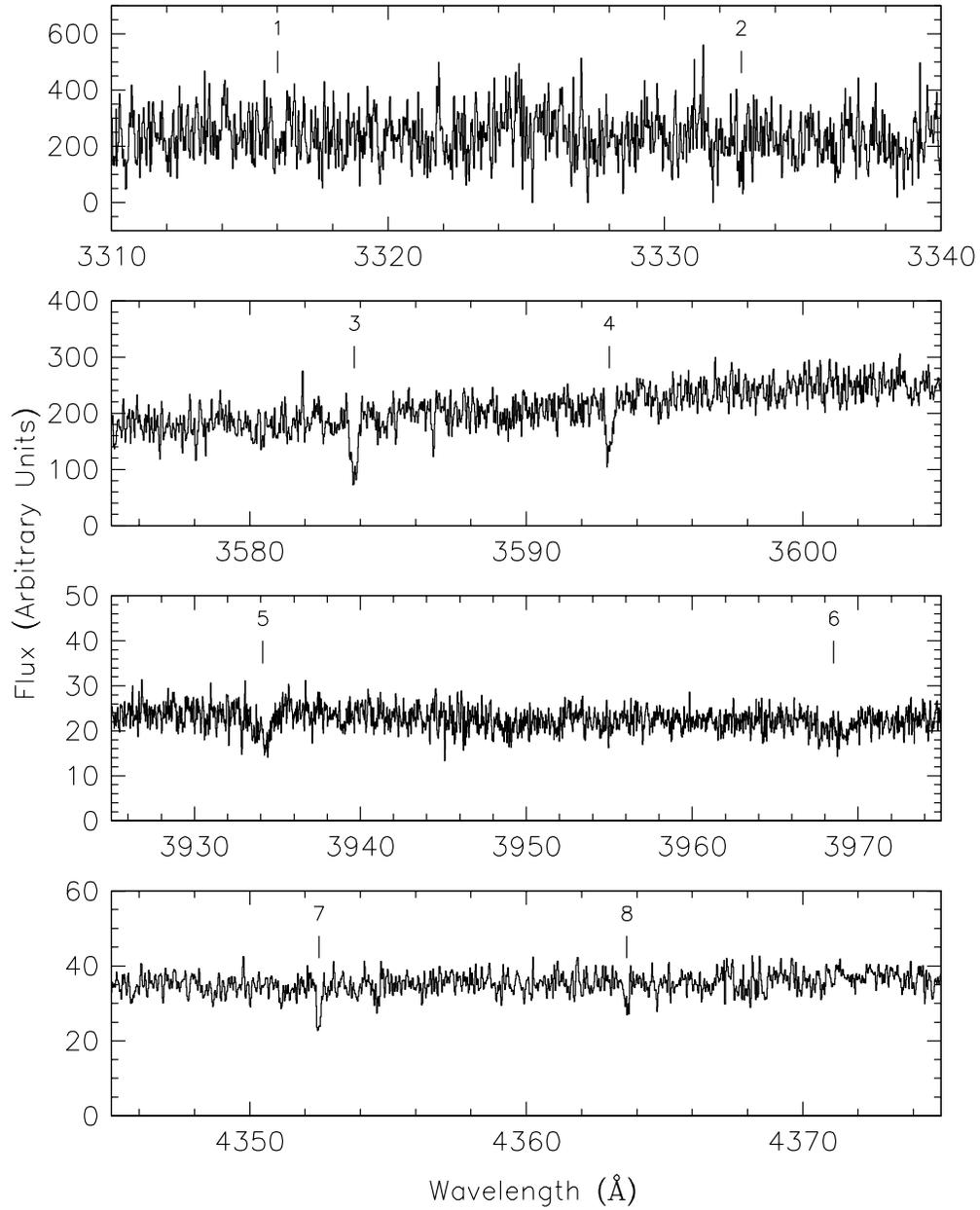}
\caption[fig06.eps]{Spectrum of the $z=1.326$ QSO J123414+6226,
  taken with the Keck HIRES spectrograph, in regions near measured absorption
  lines.  Absorption lines are marked with vertical lines and labeled with
  their corresponding numbers from Table\,\ref{systems}.  \label{keckfig}}
\end{figure}

\begin{figure}
\epsscale{0.8}
\plotone{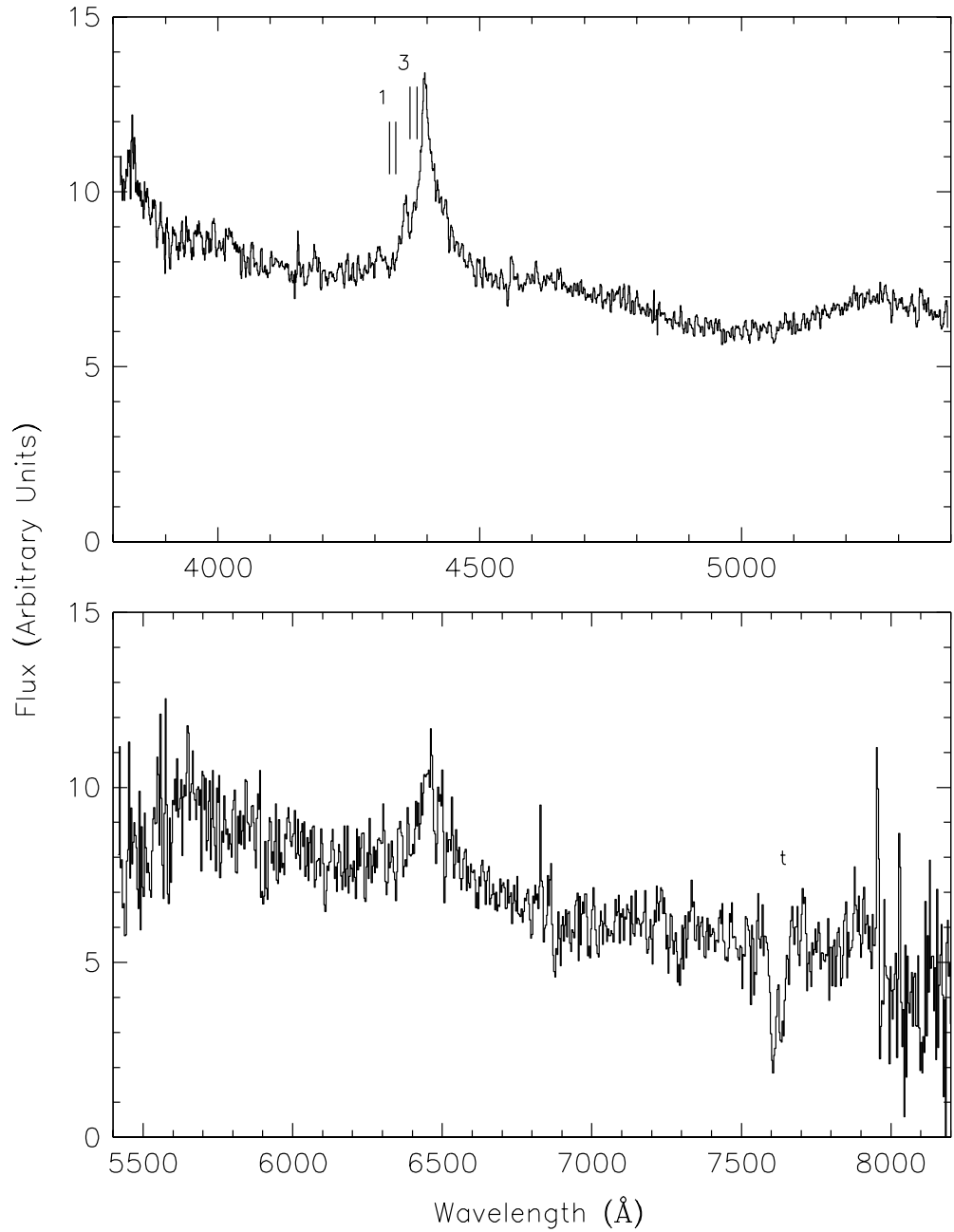}
\caption[fig07.eps]{Spectrum of the $z=1.305$ QSO J123402+6227,
  taken with the ARC/DIS and MDM/Modular spectrographs.  Absorption lines are
  marked with vertical lines and several are labeled with their corresponding
  numbers from Table\,\ref{systems}.  Telluric absorption is labeled with a
  ``t'' \label{apofig}}
\end{figure}

\begin{figure}
\epsscale{0.8}
\plotone{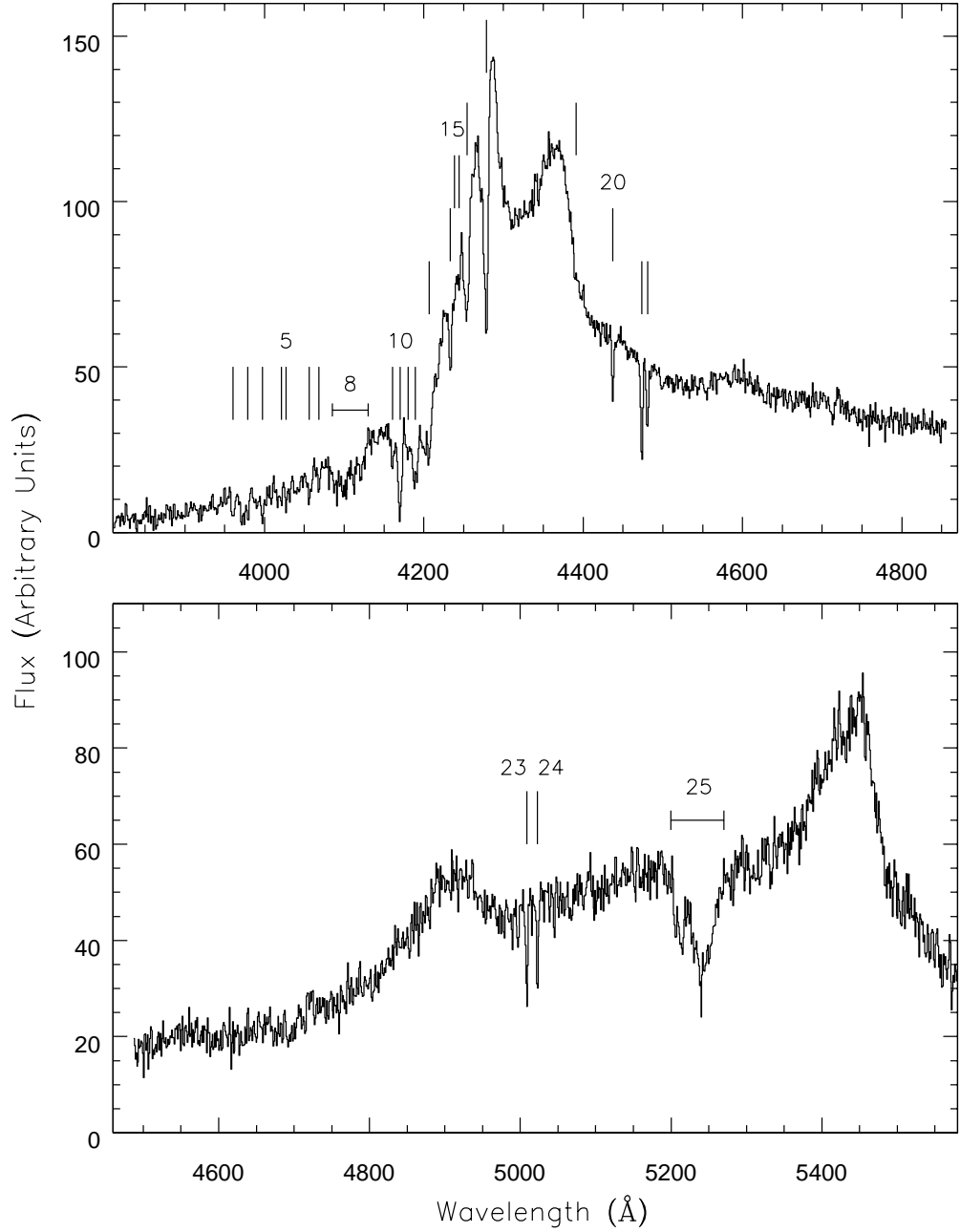}
\caption[fig08.eps]{Spectrum of the $z=2.518$ QSO J123637+6158,
  taken with the MDM/Modular spectrograph.  Absorption lines are marked with
  vertical lines and several are labeled with their corresponding numbers
  from Table\,\ref{systems}. \label{mdmfig}}
\end{figure}

The three QSO spectra were searched for absorption lines using the methods
described by \citet{vandenberk1999}.  Briefly, a continuum is fit to 
the spectrum, the flux spectrum and error arrays are normalized by the fit,
then convolved with a normalized line-spread-function profile to produce an
``equivalent-width'' array.  Absorption features having a significance level
above 3$\sigma$ were flagged, then measured by fitting Gaussian profiles
which yield observed line centers, equivalent widths, and their associated
uncertainties.  The lines were identified with ionic transitions and 
redshifts based upon the line positions, strengths, and presence of 
corroborating lines.  In the final line list, only lines with a significance
level greater then $4.5\sigma$ were kept, unless the line could be identified
with a transition occurring in a system identified with more significant lines.
The absorption lines are listed in Table~\ref{systems} and marked on the
QSO spectra plots in Figs.\,\ref{keckfig}--\ref{mdmfig}.

Not counting \lya forest lines, a Milky Way ISM system,  and a BAL system,
we have identified 5 heavy-element absorption line systems in the
three QSO spectra -- two each in J123414+6226 and J123637+6158, and one in
J123402+6227.   The systems in the Keck spectrum of J123414+6226 are at
$z=0.28159$ and $z=0.55649$, and both are \mgii doublet systems.  Both systems 
would be classified as ``weak'' since their equivalent widths are less than
$0.3${\AA} \citep{churchill1999}.
The system in the ARC/MDM spectrum of J123402+6227 is at a redshift of 
$z=0.5621$.  The line widths are somewhat uncertain since they lie on the
blueward edge of the QSO C\,{\sc iii}]\,$\lambda 1909$ emission line, but
the line centers and relative equivalent widths are consistent with a
\mgii doublet.  There is another possible \mgii doublet in this spectrum
at $z= 0.5478$, but we list it only as a candidate, since the doublet ratios
are inconsistent, and one line falls below our $4.5\sigma$ completeness limit.
No significant lines were detected in the red spectrum of the QSO, but the
$4.5\sigma$ lower equivalent width limit of this spectrum, $\sim 1.3${\AA}, is
relatively insensitive.
The two systems in the MDM spectrum of J123637+6158 are at $z=0.7913$ and
$z=1.8895$, and are identified by a \mgii and \civ doublet respectively.
The spectrum also shows a rich \lya forest ranging from $2.15 < z < 2.52$,
and a broad absorption line system near $z = 2.38$, seen in both \lya
and \civ absorption. Since the BAL phenomena is likely to be unrelated to
intervening galaxies \citep{turnshek1984}, we have not included it in the
analysis of \S\,\ref{discsec}. 

\section{Comparison With the HDF Galaxy Redshift Distribution \label{discsec}}
There are so far about $300$ published galaxy redshifts towards the HDF,
which come mainly from the surveys of 
\citet{cohen1996a,steidel1996,lowenthal1997,guzman1997,phillips1997},
and \citet{hogg1998}.
The galaxy redshifts measured towards the HDF lie within two redshift ranges,
$0\lesssim z \lesssim 1.3$ and $2.9 \lesssim z \lesssim 3.6$.  There are
few measured redshifts between these ranges, due to the lack of prominent
spectral features observable at optical wavelengths.  QSOs are observable
over this entire redshift range, but our highest confirmed QSO has a
redshift of $z=2.58$.  Absorption systems are detectable at wavelengths
above the atmospheric cutoff at $z \gtrsim 0.15$ for \mgii doublets.
Thus the distributions of galaxies, QSOs, and absorbers can be compared
within overlapping redshift ranges.

\begin{figure}
\epsscale{0.8}
\plotone{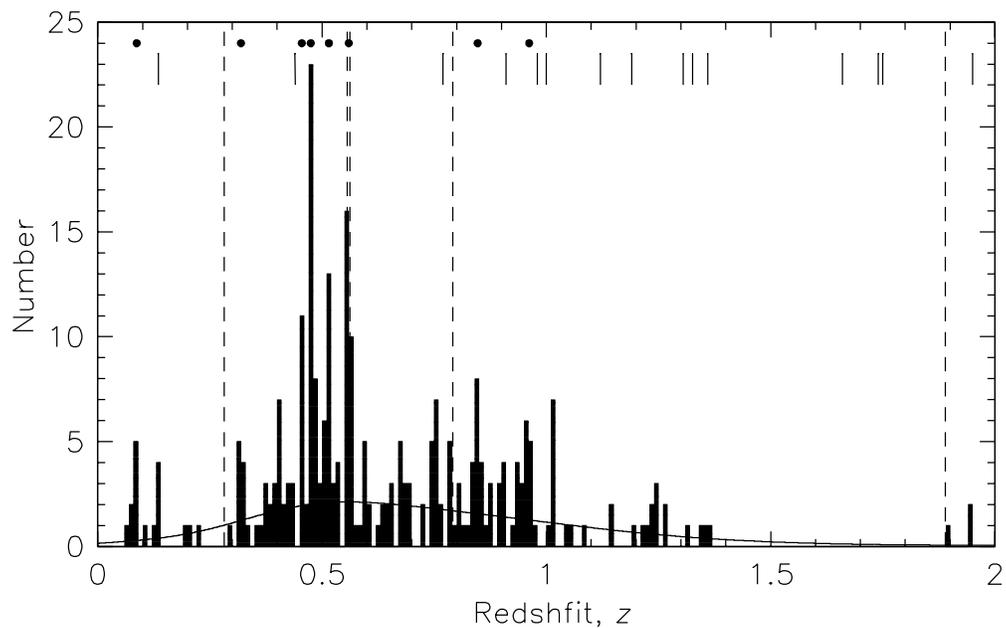}
\caption[fig09.eps]{Redshift distribution of galaxies towards
  the HDF (solid histogram) in bins of $\Delta z = 0.01$.  The redshifts of
  significant peaks (as defined in the text) are marked with dots.  The
  redshifts of QSOs are shown with solid vertical lines, and the absorption
  line system redshifts are shown with long dashed lines.  The solid curve is
  the galaxy redshift distribution smoothed by a Gaussian kernel with a width
  of $\sigma = 15000$km/s.  \label{hdfz}}
\end{figure}

The redshift distribution of the galaxies towards the HDF up to $z=2$ is
shown in Fig.\,\ref{hdfz}, and the redshifts of the individual QSOs and
absorbers are superimposed.  The galaxy distribution is characterized by sharp
peaks which contain most of the galaxies.
We have defined redshift peaks in a manner similar to \citet{cohen1996b}.
The statistical significance parameter, $X_{max}$, is defined as the 
maximum absolute number of standard deviations the number count in a peak
lies from the mean count, found after varying the count histogram bin sizes
and locations \citep{cohen1996b}.  A group of galaxies is considered a
peak if the group contains at least 5 galaxies, and has an $X_{max}\geq5$.
At least 8 distinct peaks are identified this way at redshifts 
0.087, 0.319, 0.455, 0.475, 0.515, 0.559, 0.847, and 0.962, which are
marked by dots on Fig.\,\ref{hdfz}.  This list includes 5 of the 6 peaks
identified by \citet{cohen1996b}, excluding the peak at 0.679 which we find
has 7 members but an $X_{max}$ of only $3.9$.  Many less significant peaks
may also be present.  The velocity widths of the peaks are typically
$\sigma_{v} \approx 300$km/s.  If these structures are the precursors
to walls seen in the local universe, as suggested by \citet{cohen1996b}, some
of the QSOs and absorbers in the surrounding volume are likely to be 
contained within these structures.  Of the 19 QSOs and 1 AGN, few appear to
be coincident with any of the strong galaxy peaks, but several have redshifts
close to possible smaller galaxy groups (Fig.\,\ref {hdfz}).  The measured
redshifts of QSOs can vary by over 1000km/s depending on what emission lines
are used and how they are measured \citep[\eg][]{tytler1992}.
For this reason, QSOs are not ideal for tracing structure on scales less than a 
few hundred km/s, and any matches between galaxy peak and QSO redshifts
would be uncertain.

Redshifts for absorption line systems, on the other hand, can be measured very
accurately, even with relatively low-resolution absorption spectra.  Of the
four absorbers which lie in the well-sampled galaxy redshift range
($z\la 1.3$), two have redshifts coinciding with the second most
populated peak in the galaxy distribution at $z\approx 0.559$, one
system at $z=0.7913$ lies near a possible weaker galaxy group, and one
at $z=0.2816$ does not appear to lie near any galaxy feature.
The eight galaxy peaks occupy a total velocity path of about $4800$km/s between
$0 < z \leq 1.3$ (assuming $600$km/s per peak) or about $2\%$ of the total
velocity path, so the random binomial probability of finding two or more out
of 4 absorption systems in any of the peaks is about $0.2\%$.
Thus it is reasonable to assume that at least some of the absorption line
systems are physically related to the peaks in the galaxy distribution.

If the absorbers at $z=0.559$ are parts of the same structure that contains
the galaxies, then the galaxy structure extends at least as far as the
HDF and absorber transverse separations.  For an $\Omega_{m}=1$,
$\Omega_{\Lambda}=0$ ($\Omega_{m}=0.3$, $\Omega_{\Lambda}=0.7$) universe, the
comoving transverse separations of the QSO absorbers and HDF are
$7.9$ ($9.6$) and $8.5$ ($10.4$) $h^{-1}$Mpc.  The inclination angle of
a hypothesized sheet containing the galaxies and absorbers to the line of
sight would likely be less than $30$ degrees, given that each absorber is 
about one velocity dispersion width ($\approx 460$km/s) from
the mean redshift of the galaxy peak, and on opposite sides.  Even for fairly
large inclination angles, we would expect absorption members of the sheet to
lie close to the redshift of the galaxy peak at this transverse separation,
since the velocity width of the peak translated into a comoving width is
$\approx 3.7$ ($5.3$) $h^{-1}$Mpc at $z=0.559$. At this preliminary stage,
the combined galaxy and absorption data are consistent with the suggestion by
\citet{cohen1996a, cohen1996b} that this structure and those containing 
other galaxy peaks are parts of the precursors to present-day superclusters
or walls.  The lower limit on the transverse size of the structure at
$z=0.559$ is about twice the radial extent, but a denser and wider
absorption study is needed to definitively test for a filamentary or
sheet-like geometry.

There is a strong correlation between the presence of a \mgii absorption line
system and a luminous galaxy in close physical proximity \citep{bergeron1991,
steidel1994,guillemin1997}.
It is therefore probably not surprising to find a number of these systems
near concentrations of luminous galaxies in redshift space.  Our preliminary
result from the three QSO sightlines demonstrates the utility of using
heavy-element QSO absorption line systems as complementary probes of 
large-scale structure at high redshift.  Absorption spectroscopy of the
remaining QSOs in our sample, and those of the slightly wider survey of
LPIF, would likely yield an order of magnitude more absorption line systems
towards the HDF.  Such a sample could show, for example, whether the 
absorbers and galaxies occupy the redshift peaks at the same frequency,
how far the galaxy structures extend in three dimensions, and how the 
absorbers and galaxies are biased relative to one another.

\section{Summary \label{summary}}
We have begun a survey to identify QSOs and absorption line systems in
a $45 \times 45$ square arcmin area surrounding the Hubble Deep Field.
So far $19$ QSOs have been identified within our survey area to a limiting
magnitude of $B\sim21$, and over 30 UVX and high-redshift QSO candidates
remain.  We have obtained absorption line spectra for three of the 
brighter QSOs in the field, which have revealed at least 5 heavy-element
absorption line systems.  Of the four systems that overlap the redshift
range explored in deep galaxy redshift surveys of the HDF, two lie at
or very near one of the strongest redshift peaks in the galaxy distribution.
If the absorbers and galaxies in the peak are part of the same structure,
it extends at least $7h^{-1}$Mpc ($\Omega_{m}=1$, $\Omega_{\Lambda}=0$) in the
transverse direction at a redshift
of $z\approx 0.56$.  This supports earlier evidence from the galaxies alone
that the peaks in the galaxy distribution are parts of larger structures,
which may be the precursors to present-day superclusters or walls.

\acknowledgments
We are grateful to Inger J{\o}rgensen, Marcel Bergmann, and Gary Hill
for assistance in developing the image reduction routines.
D.E.V.B. was supported in part by the Harlan J. Smith Fellowship at the
University of Texas McDonald Observatory.


\newpage

\clearpage
\input{tab01}
\clearpage
\input{tab02}
\clearpage
\input{tab03}
\clearpage
\input{tab04}
\clearpage
\input{tab05}
\clearpage
\input{tab06}
\clearpage
\input{tab07}
\clearpage
\input{tab08}

\end{document}

%% file: tab01.tex
\begin{deluxetable}{lcc}
\tablewidth{0in}
\tablecaption{Co-added Science Images \label{imagelog}}
\tablehead{
\colhead{Filter} &
\colhead{Exposure (s)} &
\colhead{FWHM (pix)}
}
\startdata
U & 3600 & 2.53 \\
B & 4200 & 2.58 \\
V & 2700 & 2.27 \\
R & 2400 & 2.21 \\
I & 2400 & 2.36 \\ 
\enddata
\end{deluxetable}

%% file: tab02.tex
\begin{deluxetable}{lcrr}
\tablecolumns{4}
\tablewidth{0pt}
\tablecaption{QSO Selection Criteria \label{qsocut}}
\tablehead{
\colhead{} & \colhead{} & \multicolumn{2}{c}{Number of Candidates} \\
\colhead{Color Plane} &
\colhead{Color Limits} &
\colhead{``Bright''} &
\colhead{``Faint''}
}
\startdata
$(U-B)/(B-V)$ & $U-B \leq 0.3 \ {\rm or} \  B-V \leq 0.35$ & 43 & 94 \\
$(U-B)/(B-V)$ & $0.0 \leq U-B \leq 1.4 \ {\rm and} \ U-B \geq 1.65(B-V)-0.58$
  & 1 & 2 \\
$(U-V)/(V-R)$ & $0.3 \leq U-V \leq 2.9 \ {\rm and} \ U-V \geq 5.78(V-R)-1.14$
  & 2 & 1 \\
$(B-V)/(V-R)$ & $0.5 \leq B-V \leq 1.9 \ {\rm and} \ B-V \geq 2.00(V-R)-0.10$
  & 4 & 8 \\
$(B-R)/(R-I)$ & $0.5 \leq B-R \leq 3.2 \ {\rm and} \ B-R \geq 3.00(V-R)-0.50$
  & 2 & (8)\tablenotemark{a} \\
\enddata
\tablenotetext{a}{These objects are already counted among those above,
since faint candidates are required to have passed at least 2 selection cuts.}
\end{deluxetable}

%% file: tab03.tex
\begin{deluxetable}{llrrlll}
\tablewidth{0pt}
\tablecaption{Unconfirmed UVX QSO Candidates \label{qsouvx}}
\tablehead{
\colhead{$\alpha$ J2000} &
\colhead{$\delta$ J2000} &
\colhead{$U-B$} &
\colhead{$B-V$} &
\colhead{$B$} &
\colhead{$R$} &
\colhead{$I$}
}
\startdata
\cutinhead{Bright UVX Candidates}
12:33:47.41 & 62:14:09.8 & -0.31 & 0.53 & 19.89 & 19.02 & 18.80 \\
12:33:57.07 & 62:34:04.9 & -0.30 & 0.57 & 19.85 & 18.90 & 18.95 \\
12:34:16.12 & 62:14:49.9 & -0.33 & 0.48 & 19.27 & 18.47 & 18.10 \\
12:34:23.70 & 61:54:44.3 & -0.57 & 0.60 & 20.33 & 19.25 & 18.87 \\
12:34:40.84 & 62:20:10.4 & -0.93 & 0.54 & 20.00 & 18.89 & 18.40 \\
12:34:51.33 & 62:26:14.1 & -0.62 & 0.13 & 20.69 & 20.46 & 59.00 \\
12:35:28.09 & 62:31:17.0 & -0.15 & 0.29 & 18.08 & 17.49 & 17.25 \\
12:35:38.50 & 62:16:44.7 & -0.38 & 0.25 & 19.82 & \nodata & \nodata \\
12:35:53.81 & 62:25:17.7 & -0.43 & 0.26 & 20.27 & 19.59 & 19.12 \\
12:36:18.72 & 61:54:09.8 & -0.77 & 0.88 & 20.19 & 18.91 & 18.37 \\
12:37:06.78 & 62:17:03.4 & -1.07 & 0.53 & 20.32 & 19.76 & 19.51 \\
12:37:53.90 & 62:19:27.2 & -0.34 & 0.42 & 20.30 & 19.58 & 19.34 \\
12:37:55.83 & 62:00:41.4 & -0.39 & 0.65 & 20.30 & 19.31 & 18.86 \\
12:38:47.27 & 62:14:03.8 & -0.34 & 0.38 & 19.87 & 19.13 & 18.99 \\
12:38:55.25 & 62:13:26.9 & -0.24 & 0.33 & 20.19 & 19.52 & 19.17 \\
12:39:23.19 & 62:13:12.5 & -0.39 & 0.59 & 19.32 & 18.34 & 17.95 \\
12:39:26.22 & 62:34:05.6 & -0.77 & 0.12 & 20.86 & 20.23 & 19.64 \\
12:39:31.76 & 62:11:48.2 & -0.39 & 0.48 & 20.47 & 19.62 & 19.49 \\
\cutinhead{Faint UVX Candidates}
12:33:47.89 & 61:53:40.1 & -1.17 &  0.68 & 22.15 & 20.93 & 20.53 \\
12:33:51.71 & 62:26:58.2 & -0.46 &  0.32 & 22.06 & 21.55 & \nodata \\
12:33:51.87 & 61:55:30.2 & -0.50 &  0.81 & 21.48 & 20.37 & 19.83 \\
12:34:03.50 & 62:30:39.7 & \nodata &  0.06 & 22.19 & 21.62 & 21.38 \\
12:34:06.62 & 62:07:45.9 & -0.03 &  0.29 & 21.59 & 21.16 & 20.20 \\
12:34:06.64 & 61:56:06.9 & -0.41 &  1.18 & 21.11 & 18.81 & 17.32 \\
12:34:10.38 & 62:02:59.4 & -0.25 & -0.04 & 21.61 & 20.83 & 20.18 \\
12:34:15.24 & 61:55:01.2 & -0.55 &  0.39 & 21.46 & 20.56 & 19.95 \\
12:34:21.70 & 62:17:02.5 & \nodata &  0.03 & 22.46 & 21.26 & 20.61 \\
12:34:33.24 & 62:15:15.0 & -1.03 &  0.26 & 21.16 & 20.70 & 20.57 \\
12:34:33.75 & 61:53:11.9 & -0.71 &  0.28 & 21.52 & 20.52 & 19.77 \\
12:34:57.65 & 61:54:37.7 & -0.56 &  0.99 & 21.71 & 20.41 & 19.78 \\
12:35:05.83 & 61:52:44.3 & -0.99 &  0.19 & 21.46 & 21.72 & 20.82 \\
12:35:08.07 & 61:52:39.3 & \nodata & -0.16 & 21.74 & 20.62 & 20.24 \\
12:35:08.21 & 61:53:55.3 & -0.58 &  0.76 & 22.11 & 20.29 & 20.12 \\
12:35:19.72 & 62:10:37.9 & -0.31 &  0.46 & 20.45 & 19.59 & 19.18 \\
12:35:20.08 & 61:54:40.8 & \nodata &  0.16 & 21.82 & 21.22 & 20.43 \\
12:35:35.45 & 61:52:52.2 & \nodata &  0.16 & 21.87 & 20.68 & 19.70 \\
12:35:43.08 & 62:08:34.8 & -0.22 &  0.23 & 21.55 & 20.47 & 19.98 \\
12:35:43.55 & 62:35:24.7 & -0.74 & \nodata & 21.76 & \nodata & \nodata \\
12:35:47.16 & 62:23:43.5 & -0.75 &  0.59 & 20.86 & 19.73 & 19.41 \\
12:35:49.43 & 62:29:20.4 & \nodata &  0.22 & 21.33 & 20.79 & 20.36 \\
12:35:55.58 & 62:01:06.3 & -1.50 &  0.58 & 22.62 & 20.89 & 20.33 \\
12:36:01.66 & 61:56:19.0 & -0.81 &  0.93 & 21.82 & 20.19 & 19.51 \\
12:36:04.39 & 62:00:55.1 & \nodata & -0.33 & 21.76 & 21.39 & \nodata \\
12:36:09.64 & 61:54:13.6 & \nodata & -0.08 & 22.18 & 21.15 & 20.06 \\
12:36:12.12 & 62:19:41.6 & -0.66 & -0.05 & 21.13 & 21.00 & 20.50 \\
12:36:16.42 & 61:51:16.7 & -0.42 &  0.42 & 20.57 & 19.75 & 19.51 \\
12:36:18.72 & 61:52:57.4 & -0.65 &  0.25 & 22.27 & 20.88 & 20.16 \\
12:36:19.83 & 62:22:20.3 & \nodata &  0.33 & 21.51 & 20.51 & 20.05 \\
12:36:28.15 & 62:14:33.5 & \nodata &  0.07 & 22.39 & 21.30 & 21.06 \\
12:36:41.88 & 61:54:45.4 & -0.69 &  0.39 & 21.80 & 20.87 & 20.25 \\
12:36:46.11 & 62:27:54.2 & -1.34 &  0.95 & 21.98 & 21.05 & 20.32 \\
12:36:49.49 & 62:29:34.2 & -0.45 &  0.26 & 20.95 & 20.81 & 20.60 \\
12:37:01.84 & 62:00:44.3 & -0.37 &  0.26 & 20.85 & 20.43 & 19.72 \\
12:37:03.92 & 61:53:56.1 & -0.82 &  1.27 & 21.26 & 19.19 & 18.60 \\
12:37:07.36 & 61:59:46.4 & -1.68 &  0.86 & 21.94 & 20.58 & 20.40 \\
12:37:07.43 & 61:54:17.7 & -0.51 &  0.93 & 21.68 & 20.55 & 20.22 \\
12:37:14.41 & 62:17:54.7 & -0.66 &  0.73 & 21.20 & 20.08 & 19.82 \\
12:37:17.27 & 61:56:23.5 & -0.63 &  0.70 & 21.10 & 20.00 & 19.55 \\
12:37:19.83 & 62:28:36.2 & -1.33 &  0.69 & 22.53 & 21.81 & \nodata \\
12:37:19.88 & 61:55:34.9 & \nodata &  0.34 & 21.91 & 21.07 & 20.76 \\
12:37:26.31 & 61:58:18.1 & -0.85 &  0.85 & 21.47 & 19.93 & 18.83 \\
12:37:30.84 & 62:02:22.1 & -0.90 &  0.54 & 21.06 & 20.34 & 19.63 \\
12:37:34.85 & 61:57:10.9 & \nodata &  0.28 & 21.74 & 20.72 & 20.29 \\
12:37:36.22 & 61:58:35.1 & \nodata &  0.15 & 22.19 & 21.41 & \nodata \\
12:37:49.67 & 61:54:09.2 & -1.53 &  0.59 & 22.63 & 20.90 & 20.68 \\
12:38:01.44 & 62:20:20.1 & -0.43 &  0.68 & 21.00 & 20.00 & 19.63 \\
12:38:02.56 & 62:10:44.0 & -0.73 &  0.97 & 22.04 & 20.53 & 19.82 \\
12:38:03.32 & 62:25:30.1 & -0.41 &  0.97 & 21.01 & 19.45 & 18.89 \\
12:38:06.82 & 62:06:57.5 & -0.56 &  0.60 & 20.89 & 19.75 & 19.35 \\
12:38:07.64 & 62:20:43.8 & \nodata &  0.14 & 21.46 & 21.06 & 20.64 \\
12:38:07.93 & 62:29:50.4 & \nodata &  0.15 & 22.22 & 21.54 & \nodata \\
12:38:08.75 & 62:08:36.0 & -0.40 &  0.49 & 20.69 & 19.92 & 19.43 \\
12:38:15.87 & 62:30:15.9 & -1.30 & -0.17 & 21.47 & 21.11 & 21.11 \\
12:38:19.15 & 62:32:45.7 & -1.19 &  0.29 & 21.69 & 21.16 & 20.89 \\
12:38:22.61 & 62:24:01.7 & \nodata &  0.12 & 21.89 & 20.91 & 21.00 \\
12:38:25.06 & 61:52:38.3 & -0.78 &  0.49 & 21.18 & 21.04 & 20.84 \\
12:38:29.05 & 61:51:22.8 & -0.40 &  0.88 & 21.89 & 20.52 & 20.51 \\
12:38:29.35 & 62:10:20.7 & -1.14 &  1.27 & 22.00 & 20.00 & 19.29 \\
12:38:33.54 & 62:03:52.6 & -1.05 & -0.19 & 21.22 & 21.09 & 20.51 \\
12:38:38.58 & 62:04:43.7 & -1.37 &  1.66 & 22.37 & 20.84 & 20.60 \\
12:38:43.98 & 62:18:23.5 & -0.32 &  0.86 & 21.91 & 20.85 & 20.50 \\
12:38:46.73 & 62:35:38.2 & -2.67 &  1.19 & 22.63 & 21.31 & \nodata \\
12:38:54.83 & 62:33:55.8 & -1.70 &  0.68 & 22.40 & 21.66 & 21.69 \\
12:39:02.36 & 62:20:26.7 & -1.04 &  0.48 & 21.53 & 20.58 & 20.29 \\
12:39:04.74 & 61:57:50.9 & -0.38 &  0.86 & 21.48 & 20.07 & 19.86 \\
12:39:13.02 & 62:08:53.9 & \nodata &  0.30 & 22.22 & 20.78 & 20.43 \\
12:39:13.59 & 61:52:45.3 & \nodata & -0.93 & 21.56 & 21.50 & 21.23 \\
12:39:14.35 & 62:09:57.9 & -0.40 &  0.68 & 21.54 & 20.73 & 20.44 \\
12:39:18.63 & 61:59:40.9 & \nodata &  0.22 & 22.02 & 20.85 & 20.40 \\
12:39:20.33 & 61:58:39.6 & -1.01 &  0.21 & 21.54 & 21.01 & 20.04 \\
12:39:21.67 & 61:52:27.3 & -1.12 &  0.39 & 21.34 & 20.78 & 20.67 \\
12:39:27.95 & 61:50:53.2 & -1.08 &  0.48 & 21.55 & 20.78 & 19.98 \\
12:39:31.52 & 62:17:48.6 & -0.73 &  0.71 & 21.82 & 20.71 & 20.40 \\
12:39:36.67 & 62:30:51.9 & \nodata & -0.02 & 21.77 & 20.57 & 20.01 \\
12:39:47.75 & 62:03:38.5 & \nodata &  0.26 & 22.21 & 20.57 & 19.98 \\
12:39:47.95 & 62:01:42.3 & -0.74 &  0.99 & 22.19 & 20.62 & 20.44 \\
12:39:57.57 & 62:00:08.5 & \nodata &  0.28 & 21.44 & 20.78 & 20.24 \\
12:39:58.83 & 62:13:31.0 & -1.44 &  1.07 & 22.38 & 20.75 & 20.23 \\
12:40:01.90 & 62:08:42.2 & -0.33 &  0.38 & 20.96 & 20.40 & 19.48 \\
12:40:09.17 & 61:53:32.5 & \nodata &  0.19 & 22.13 & 21.01 & 20.48 \\
12:40:16.28 & 61:59:22.5 & \nodata & -2.21 & 19.32 & \nodata & \nodata \\
12:40:18.15 & 62:17:46.7 & \nodata & -1.59 & 19.76 & \nodata & \nodata \\
\enddata
\end{deluxetable}

%% file: tab04.tex
\begin{deluxetable}{llrrlll}
\tablewidth{0pt}
\tablecaption{Unconfirmed High-$z$ QSO Candidates \label{qsohiz}}
\tablehead{
\colhead{$\alpha$ J2000} &
\colhead{$\delta$ J2000} &
\colhead{$U$} &
\colhead{$B$} &
\colhead{$V$} &
\colhead{$R$} &
\colhead{$I$}
}
\startdata
\cutinhead{Bright High-$z$ Candidates}
12:39:10.94 & 62:34:42.4 & 20.04 & 19.92 & 19.52 & 19.14 & 18.90 \\
12:37:11.36 & 62:24:27.1 & 19.15 & 19.28 & 18.58 & 18.30 & 17.91 \\
12:37:23.75 & 62:15:44.4 & 19.76 & 19.83 & 19.46 & 19.26 & 18.94 \\
12:37:48.59 & 62:19:49.0 & 20.04 & 20.29 & 19.60 & 19.39 & 19.00 \\
12:36:10.18 & 61:56:08.3 & \nodata & 20.78 & 19.60 & 19.66 & 19.29 \\
12:39:30.42 & 61:54:33.5 & 20.34 & 21.00 & 19.66 & 19.16 & 19.45 \\
12:39:37.32 & 62:18:00.7 & \nodata & 21.29 & 20.12 & 19.70 & 19.15 \\
12:39:46.04 & 62:20:00.6 & 20.17 & 20.75 & 19.27 & 18.65 & 17.99 \\
12:39:58.35 & 61:52:27.0 & 20.65 & 20.92 & 19.49 & 18.87 & 18.24 \\
12:34:33.19 & 62:34:41.0 & 20.38 & 21.23 & 20.52 & 19.80 & 19.59 \\
12:35:23.23 & 62:31:35.6 & 20.18 & 20.60 & 19.42 & 18.85 & 18.52 \\
12:35:42.04 & 62:02:01.1 & 20.62 & 20.35 & 19.41 & 18.96 & 18.67 \\
\cutinhead{Faint High-$z$ Candidates}
12:33:50.96 & 61:55:59.9 & \nodata & 22.89 & 21.42 & 20.85 & 20.51 \\
12:35:12.73 & 61:52:16.7 & \nodata & 22.84 & 21.42 & 21.23 & 20.99 \\
12:36:07.35 & 61:53:35.6 & \nodata & 22.31 & 21.04 & 20.74 & 20.40 \\
12:36:11.49 & 62:32:12.0 & 20.91 & 20.71 & 20.24 & 20.16 & 19.75 \\
12:36:14.24 & 61:51:53.9 & \nodata & 22.80 & 21.17 & 20.88 & 20.53 \\
12:38:02.00 & 62:15:20.5 & \nodata & 22.05 & 20.55 & 20.14 & 19.77 \\
12:38:16.18 & 62:33:51.8 & 21.53 & 21.34 & 20.23 & 19.82 & 19.33 \\
12:38:21.67 & 61:56:30.6 & \nodata & 22.47 & 21.75 & 21.61 & 21.63 \\
12:38:37.03 & 61:51:27.3 & \nodata & 22.83 & 20.70 & 19.80 & 19.08 \\
12:39:32.98 & 62:32:39.6 & \nodata & 22.54 & 21.24 & 21.04 & 20.79 \\
\enddata
\end{deluxetable}

%% file: tab05.tex
\begin{deluxetable}{lllllllllllllllllr}
\tabletypesize{\scriptsize}
\tablewidth{0pt}
\tablecaption{Spectroscopically Confirmed QSOs and Galaxies \label{qsoid}}
\tablehead{
\colhead{ID} &
\colhead{$\alpha$ J2000} &
\colhead{$\delta$ J2000} &
\colhead{$U$} &
\colhead{$\sigma_{U}$} &
\colhead{$B$} &
\colhead{$\sigma_{B}$} &
\colhead{$V$} &
\colhead{$\sigma_{V}$} &
\colhead{$R$} &
\colhead{$\sigma_{R}$} &
\colhead{$I$} &
\colhead{$\sigma_{I}$} &
\colhead{$z_{em}$}
}
\startdata
\cutinhead{QSOs}
J123401+6233 & 12:34:01.04 & 62:33:15.6 & 19.48 & 0.07 & 20.36 & 0.06 & 19.86
 & 0.05 & 19.77 & 0.04 & 19.44 & 0.08 & 1.12\tablenotemark{a} \\
J123402+6227 & 12:34:02.49 & 62:27:52.5 & 18.50 & 0.04 & 19.51 & 0.04 & 19.19
 & 0.03 & 18.77 & 0.02 & 18.47 & 0.03 & 1.305 \\
J123411+6158 & 12:34:11.71 & 61:58:32.8 & 18.34 & 0.04 & 19.39 & 0.03 & 18.77
 & 0.02 & 18.48 & 0.02 & 17.89 & 0.02 & 1.95\tablenotemark{a} \\
J123414+6226 & 12:34:14.80 & 62:26:40.2 & 16.62 & 0.01 & 17.63 & 0.01 & 17.43
 & 0.01 & 17.05 & 0.01 & 16.73 & 0.01 & 1.326 \\
J123426+6154 & 12:34:26.64 & 61:54:32.4 & 18.90 & 0.05 & 19.36 & 0.03 & 19.18
 & 0.03 & 19.14 & 0.02 & 18.74 & 0.03 & 2.514 \\
J123428+6208 & 12:34:28.24 & 62:08:23.8 & 20.12 & 0.11 & 21.03 & 0.09 & 20.96
 & 0.10 & 20.47 & 0.07 & 20.59 & 0.12 & 1.355 \\
J123428+6206 & 12:34:28.41 & 62:06:32.1 & 19.79 & 0.09 & 20.47 & 0.07 & 20.02
  & 0.05 & 19.72 & 0.04 & 19.05 & 0.05 & 1.793 \\
J123512+6150 & 12:35:12.97 & 61:50:57.1 & 19.97 & 0.10 & 20.85 & 0.08 & 20.47
 & 0.06 & 20.29 & 0.05 & 19.70 & 0.08 & 0.98\tablenotemark{a} \\
J123610+6204 & 12:36:10.24 & 62:04:35.3 & 19.01 & 0.06 & 19.86 & 0.05 & 19.93
 & 0.05 & 19.35 & 0.04 & 19.30 & 0.05 & 1.74\tablenotemark{a} \\
J123622+6215 & 12:36:22.89 & 62:15:27.4 & 20.50 & 0.13 & 20.50 & 0.08 & 20.42
 & 0.07 & 20.34 & 0.06 & 20.24 & 0.10 & 2.58\tablenotemark{a} \\
J123637+6158 & 12:36:37.45 & 61:58:15.6 & 18.75 & 0.04 & 18.95 & 0.03 & 18.89
 & 0.02 & 18.62 & 0.02 & 18.22 & 0.02 & 2.518 \\
J123715+6203 & 12:37:15.96 & 62:03:24.5 & 19.16 & 0.06 & 20.18 & 0.05 & 20.04
 & 0.05 & 19.69 & 0.04 & 19.10 & 0.05 & 2.05\tablenotemark{a} \\
J123859+6211 & 12:37:59.51 & 62:11:03.4 & 17.92 & 0.03 & 18.77 & 0.02 & 18.45
 & 0.02 & 18.31 & 0.01 & 18.14 & 0.02 & 0.910 \\
J123800+6213 & 12:38:00.85 & 62:13:36.8 & 18.31 & 0.04 & 19.16 & 0.03 & 18.87
 & 0.02 & 18.41 & 0.02 & 17.93 & 0.02 & 0.44\tablenotemark{a} \\
J123811+6227 & 12:38:11.99 & 62:27:27.5 & 20.07 & 0.12 & 20.76 & 0.08 & 20.23
 & 0.07 & 20.15 & 0.05 & 19.49 & 0.06 & 0.77\tablenotemark{a} \\
J123815+6224 & 12:38:15.46 & 62:24:40.7 & 20.25 & 0.18 & 21.33 & 0.11 & 20.86
 & 0.09 & 20.39 & 0.06 & 19.80 & 0.08 & 1.75\tablenotemark{a} \\
J123816+6202 & 12:38:16.06 & 62:02:09.2 & 18.27 & 0.03 & 19.18 & 0.03 & 18.80
 & 0.02 & 18.59 & 0.02 & 18.40 & 0.03 & 1.002 \\
J123931+6206 & 12:39:31.44 & 62:06:20.1 & 18.60 & 0.04 & 19.54 & 0.04 & 19.25
 & 0.03 & 18.89 & 0.02 & 18.66 & 0.03 & 1.19\tablenotemark{a} \\
J123933+6233 & 12:39:33.93 & 62:33:21.8 & 19.32 & 0.07 & 19.99 & 0.05 & 19.77
 & 0.04 & 19.38 & 0.03 & 18.61 & 0.03 & 1.66\tablenotemark{a} \\
\cutinhead{Galaxies}
J123429+6218 & 12:34:29.88 & 62:18:06.9 & 17.17 & 0.02 & 17.75 & 0.01 & 17.16
 & 0.01 & 16.76 & 0.01 & 16.18 & 0.01 & 0.135 \\
J123519+6200 & 12:35:19.09 & 62:00:37.2 & 20.13 & 0.13 & 20.87 & 0.08 & 19.96
 & 0.05 & 19.48 & 0.03 & 18.90 & 0.04 & 0.395 \\
J123526+6154 & 12:35:26.13 & 61:54:36.7 & 20.66 & 0.21 & 21.40 & 0.14 & 20.70
 & 0.07 & 19.80 & 0.04 & 18.61 & 0.03 & 0.250 \\
J123811+6222 & 12:38:11.83 & 62:22:40.8 & 19.89 & 0.10 & 20.49 & 0.06 & 19.95
 & 0.05 & 19.55 & 0.04 & 19.11 & 0.05 & 0.232\tablenotemark{a} \\
\enddata
\tablenotetext{a}{Spectroscopic identification and redshift from LPIF.}
\end{deluxetable}

%% file: tab06.tex
\begin{deluxetable}{lllllllllllllllllr}
\scriptsize
\tablewidth{0pt}
\tablecaption{Spectroscopically Identified Stars \label{stars}}
\tablehead{
\colhead{$\alpha$ J2000} &
\colhead{$\delta$ J2000} &
\colhead{$U$} &
\colhead{$B$} &
\colhead{$V$} &
\colhead{$R$} &
\colhead{$I$}
}
\startdata
12:34:23.89 & 62:16:36.1 & 19.86 & 20.00 & 19.71 & 19.45 & 19.23 \\
12:34:29.42 & 62:04:33.0 & 99.00 & 22.01 & 20.12 & 19.31 & 18.22 \\
12:34:40.91 & 62:29:34.6 & 20.00 & 20.46 & 20.14 & 69.00 & 59.00 \\
12:35:04.47 & 62:05:18.7 & 19.23 & 19.92 & 19.78 & 19.57 & 19.41 \\
12:35:09.56 & 62:32:53.1 & 20.13 & 20.50 & 20.00 & 19.82 & 19.43 \\
12:35:28.50 & 62:32:29.4 & 20.30 & 20.91 & 20.50 & 20.36 & 19.76 \\
12:35:44.29 & 62:34:15.2 & 18.46 & 19.42 & 19.51 & 19.55 & 19.51 \\
12:36:03.01 & 62:13:38.1 & 19.86 & 20.25 & 19.81 & 19.62 & 19.20 \\
12:36:25.29 & 62:34:22.5 & 19.52 & 19.53 & 19.35 & 19.21 & 19.22 \\
12:36:45.40 & 62:12:14.7 & 19.63 & 20.88 & 20.69 & 20.86 & 20.60 \\
12:38:04.41 & 62:10:16.8 & 20.15 & 20.07 & 19.94 & 19.72 & 19.57 \\
12:39:10.88 & 62:02:18.7 & 15.56 & 16.48 & 16.74 & 16.85 & 16.97 \\
12:39:52.14 & 61:57:04.5 & 99.00 & 22.59 & 21.33 & 20.23 & 19.89 \\
12:39:52.17 & 61:50:56.5 & 18.23 & 19.17 & 19.12 & 19.09 & 19.11 \\
\enddata
\end{deluxetable}

%% file: tab07.tex
\begin{deluxetable}{llrllll}
\tabletypesize{\small}
\tablecaption{QSO Absorption Spectra Follow-up Observations \label{hireslog}}
\tablehead{
\colhead{Object ID} &
\colhead{UT Dates} &
\colhead{Exp.\,(s)} &
\colhead{FWHM} &
\colhead{$\lambda_{low}$({\AA})} &
\colhead{$\lambda_{high}$({\AA})} &
\colhead{Spectrograph}
}
\startdata
1234+6227 & 05/29/98 &  5400 & 8 km/s & 3104 & 4661 & Keck/HIRES \\
1234+6228 & 12/23/98 & 13500 & 5.8{\AA} & 3814 & 5393 & ARC/DIS-blue \\
          & 12/23/98 & 13500 & 6.9{\AA} & 5174 & 7947 & ARC/DIS-red \\
          & 12/27/98 &  1540 & 2.2{\AA} & 3812 & 4855 & MDM2.4m/Modular \\
1237+6158 & 12/24-26/98 & 15500 & 2.2{\AA} & 3812 & 4855 &
  MDM2.4m/Modular \\
          & 12/28-29/98 & 18000 & 2.0{\AA} & 4488 & 5587 &
  MDM2.4m/Modular \\
\enddata
\end{deluxetable}

%% file: tab08.tex
\begin{deluxetable}{rllrlr}
\tabletypesize{\small}
\tablewidth{0pt}
\tablecaption{QSO absorption line systems \label{systems}}
\tablehead{
\colhead{No.} &
\colhead{$\lambda_{obs}$ ({\AA})} &
\colhead{$W_{obs}$ ({\AA})} &
\colhead{$S/N$} &
\colhead{Identification} &
\colhead{$z_{abs}$}
}
\startdata
\cutinhead{J123402+6227, $z_{em}=1.305$}
 1 & 4327.67 $\pm$ 0.64 & 0.56 $\pm$ 0.15 & 3.6 & \mgii$\lambda2796$? &
  0.5476 \\
 2 & 4339.65 $\pm$ 0.43 & 0.75 $\pm$ 0.13 & 5.7 & \mgii$\lambda2803$? &
  0.5479 \\
 3 & 4367.24 $\pm$ 0.32 & 0.68 $\pm$ 0.10 & 6.5 & \mgii$\lambda2796$ &
  0.5618 \\
 4 & 4380.22 $\pm$ 0.99 & 0.55 $\pm$ 0.11 & 5.0 & \mgii$\lambda2803$ &
  0.5624 \\
\cutinhead{J123414+6226, $z_{em}=1.326$} 
1 & 3316.042 $\pm$ 0.070 & 0.111 $\pm$ 0.034 &  3.3 & \feii$\lambda 2586$ &
  0.28198 \\
2 & 3332.768 $\pm$ 0.070 & 0.108 $\pm$ 0.034 &  3.2 & \feii$\lambda 2600$ &
  0.28175 \\
3 & 3583.779 $\pm$ 0.012 & 0.159 $\pm$ 0.015 & 10.6 & \mgii$\lambda 2796$ &
  0.28159 \\
4 & 3592.996 $\pm$ 0.017 & 0.109 $\pm$ 0.014 &  7.8 & \mgii$\lambda 2803$ &
  0.28160 \\
5 & 3934.126 $\pm$ 0.043 & 0.263 $\pm$ 0.022 & 12.0 & \caii$\lambda 3934$ &
  -0.00017 \\
6 & 3968.847 $\pm$ 0.066 & 0.160 $\pm$ 0.021 &  7.6 & \caii$\lambda 3969$ &
  -0.00019 \\
7 & 4352.497 $\pm$ 0.009 & 0.061 $\pm$ 0.009 &  6.8 & \mgii$\lambda 2796$ &
  0.55649 \\
8 & 4363.634 $\pm$ 0.024 & 0.038 $\pm$ 0.010 &  3.8 & \mgii$\lambda 2803$ &
  0.55648 \\
\cutinhead{J123637+6158, $z_{em}=2.518$}
 1 & 3960.68 $\pm$ 0.30 & 2.38 $\pm$ 0.36 & 6.6 & \lya $\lambda 1215$ &
  2.2580 \\
 2 & 3979.20 $\pm$ 0.11 & 1.84 $\pm$ 0.37 & 4.9 & \lya $\lambda 1215$ &
  2.2733 \\
 3 & 3997.27 $\pm$ 0.07 & 2.37 $\pm$ 0.18 & 13.1 & \lya $\lambda 1215$ &
  2.2881 \\
 4 & 4021.14 $\pm$ 0.24 & 0.80 $\pm$ 0.16 & 5.0 & \lya $\lambda 1215$ &
  2.3078 \\
 5 & 4027.05 $\pm$ 0.15 & 1.16 $\pm$ 0.16 & 7.4 & \lya $\lambda 1215$ &
  2.3126 \\
   &                    &                 &     & \siv$\lambda 1393$? &
  1.8894 \\
 6 & 4055.90 $\pm$ 0.20 & 1.42 $\pm$ 0.37 & 3.9 & \lya $\lambda 1215$ &
  2.3364 \\
   &                    &                 &     & \siv$\lambda 1402$? &
  1.8914 \\
 7 & 4068.32 $\pm$ 0.04 & 1.07 $\pm$ 0.09 & 12.0 & \lya $\lambda 1215$ &
  2.3466 \\
 8 & 4085--4130         & 14.0 $\pm$ 0.80 & 17.5 & \lya BAL & 2.38\\
 9 & 4160.84 $\pm$ 0.28 & 1.56 $\pm$ 0.33 & 4.7 & \lya $\lambda 1215$ &
  2.4227 \\
10 & 4170.02 $\pm$ 0.10 & 4.53 $\pm$ 0.25 & 18.3 & \lya $\lambda 1215$ &
  2.4302 \\
11 & 4180.23 $\pm$ 0.43 & 0.67 $\pm$ 0.14 & 4.6 & \lya $\lambda 1215$ &
  2.4386 \\
12 & 4189.08 $\pm$ 0.45 & 3.43 $\pm$ 0.50 & 6.8 & \lya $\lambda 1215$ &
  2.4459 \\
13 & 4206.98 $\pm$ 0.42 & 3.42 $\pm$ 0.41 & 8.3 & \lya $\lambda 1215$ &
  2.4606 \\
14 & 4233.32 $\pm$ 0.09 & 1.04 $\pm$ 0.07 & 14.8 & \lya $\lambda 1215$ &
  2.4823 \\
15 & 4238.47 $\pm$ 0.28 & 0.20 $\pm$ 0.04 & 5.0 & \lya $\lambda 1215$ &
  2.4865 \\
16 & 4244.44 $\pm$ 0.19 & 0.49 $\pm$ 0.09 & 5.4 & \lya $\lambda 1215$ &
  2.4914 \\
17 & 4254.15 $\pm$ 0.20 & 2.66 $\pm$ 0.19 & 13.7 & \lya $\lambda 1215$ &
  2.4994 \\
18 & 4278.66 $\pm$ 0.12 & 2.46 $\pm$ 0.15 & 16.8 & \lya $\lambda 1215$ &
  2.5196 \\
19 & 4391.15 $\pm$ 0.71 & 1.31 $\pm$ 0.18 & 7.5 & unknown & \\
20 & 4437.30 $\pm$ 0.09 & 0.79 $\pm$ 0.11 & 7.1 & unknown & \\
21 & 4473.49 $\pm$ 0.04 & 1.78 $\pm$ 0.14 & 12.9 & \civ$\lambda 1548$ &
  1.8895 \\
22 & 4480.79 $\pm$ 0.05 & 1.17 $\pm$ 0.10 & 11.9 & \civ$\lambda 1550$ &
  1.8894 \\
23 & 5008.74 $\pm$ 0.08 & 1.41 $\pm$ 0.18 & 8.0 & \mgii$\lambda 2796$ &
  0.7912 \\
24 & 5022.64 $\pm$ 0.06 & 1.08 $\pm$ 0.11 & 9.9 & \mgii$\lambda 2803$ &
  0.7915 \\
25 & 5200--5270         & 16.0 $\pm$ 0.75 & 21.3 & \civ BAL & 2.38 \\
\enddata
\end{deluxetable}

%% file: VandenBerk.bbl
\newpage
\begin{thebibliography}{}
\bibitem[Adelberger et~al.(1998)]{adelberger1998} Adelberger, K. L.,
  Steidel, C. C., Giavalisco, M., Dickinson, M., Pettini, M.,
  \& Kellogg, M.  1998, \apj, 505, 18
\bibitem[Bergeron \& Boisse(1991)]{bergeron1991} Bergeron, J. \& 
  Boisse, P. 1991, \aap, 243, 344 
\bibitem[Bertin \& Arnouts(1996)]{bertin1996} Bertin, E., \& Arnouts, S.
  1996, \aaps, 117, 393
\bibitem[Bi \& Fang(1996)]{bi1996} Bi, H., \& Fang, L.  1996, \apj, 466, 614
\bibitem[Cen et~al.(1998)]{cen1998} Cen, R., Phelps, S., Miralda-Escude, J.,
  \& Ostriker, J. P. 1998, \apj, 496, 577
\bibitem[Churchill et~al.(1999)]{churchill1999} 
  Churchill, C. W., Rigby, J. R., Charlton, J. C., \& Vogt, S. S. 1999, \apjs, 
  120, 51
\bibitem[Cohen et~al.(1999)]{cohen1999} Cohen, J. G., Blandford, R.,
  Hogg, D. W., Pahre, M. A., \& Shopbell, P. L. 1999, \apj, 512, 30
\bibitem[Cohen et~al.(1996a)]{cohen1996a} Cohen, J. G., Cowie, L. L.,
  Hogg, D. W., Songaila, A., Blandford, R., Hu, E. M., \& Shopbell, P.
  1996a, \apjl, 471, L5
\bibitem[Cohen et~al.(1996b)]{cohen1996b} Cohen, J. G., Hogg, D. W.,
  Pahre, M. A., \& Blandford, R.  1996b, \apjl, 462, L9
\bibitem[Crotts(1985)]{crotts1985} Crotts, A. P. S. 1985, \apj, 298, 732
\bibitem[Crotts(1989)]{crotts1989} Crotts, A. P. S. 1989, \apj, 336, 550
\bibitem[Demia{\' n}ski \& Doroshkevich(1999)]{demia1999} Demia{\' n}ski, M., \&
  Doroshkevich, A. G. 1999, \apj, 512, 527
\bibitem[Dinshaw \& Impey(1996)]{dinshaw1996} Dinshaw, N. \&
  Impey, C. D. 1996, \apj, 458, 73
\bibitem[Elowitz et~al.(1995)]{elowitz1995} Elowitz, R. M., Green, R. F.,
  \& Impey, C.  D. 1995, \apj, 440, 458
\bibitem[Fang \& Jing(1998)]{fang1998} Fang, L., \& Jing, Y. P. 1998,
  \apjl, 502, L95
\bibitem[Foltz et~al.(1993)]{foltz1993} Foltz, C. B., Hewett, P. C.,
  Chaffee, F. H., \& Hogan, C. J. 1993, \aj, 105, 22
\bibitem[Guillemin \& Bergeron(1997)]{guillemin1997} Guillemin, P., \&
  Bergeron, J. 1997, \aap, 328, 499
\bibitem[Guzm{\'a}n et~al.(1997)]{guzman1997} Guzm{\'a}n, R., Gallego, J.,
  Koo, D. C., Phillips, A. C., Lowenthal, J. D., Faber, S. M.,
  Illingworth, G. D., \& Vogt, N. P. 1997, \apj, 489, 559
\bibitem[Hall et~al.(1996)]{hall1996} Hall, P. B., Osmer, P. S., Green, R. F.,
  Porter, A. C., \& Warren, S. J. 1996, \apj, 462, 614
\bibitem[Hogg et~al.(1998)]{hogg1998} Hogg, D. W., \etal 1998, \aj, 115, 1418
\bibitem[Horne(1986)]{horne1986} Horne, K. 1986, \pasp, 98, 609
\bibitem[Impey et~al.(1999)]{impey1999} Impey, C. D., Petry, C. E.,
  \& Flint, K. P.  1999, \apj, 524, 536
\bibitem[Irwin et~al.(1991)]{irwin1991} Irwin, M., McMahon, R. G., \& Hazard, C.
  1991, in ASP Conf. Proc. 21, The Space Distribution of Quasars,
  ed. D. Crampton (San Francisco: ASP), 117
\bibitem[Jakobsen \& Perryman(1992)]{jakobsen1992} Jakobsen, P.,
  \& Perryman, M. A. C. 1992, \apj, 392, 432
\bibitem[Koo \& Kron(1988)]{koo1988} Koo, D. C., \& Kron, R. G. 1988, \apj,
  325, 92
\bibitem[Landolt(1992)]{landolt1992} Landolt, A. U. 1992, \aj, 104, 340
\bibitem[Liu et~al.(1999)]{lpif} Liu, C., Petry, C., Impey, C., \&
  Foltz, C. 1999, AJ, in press (LPIF)
\bibitem[Lowenthal et~al.(1997)]{lowenthal1997} Lowenthal, J. D., et al. 1997,
  \apj, 481, 673
\bibitem[Newberg \& Yanny(1997)]{newberg1997} Newberg, H. J., \& Yanny, B.
  1997, \apjs, 113, 89
\bibitem[Phillips et~al.(1997)]{phillips1997} Phillips, A. C., Guzm{\'a}n, R.,
  Gallego, J., Koo, D. C., Lowenthal, J. D., Vogt, N. P., Faber, S. M.,
  \& Illingworth, G.  D. 1997, \apj, 489, 543
\bibitem[Quashnock \& Vanden Berk(1998)]{quashnock1998} Quashnock, J. M.,
  \& Vanden Berk, D. E.  1998, \apj, 500, 28
\bibitem[Steidel et~al.(1996)]{steidel1996} Steidel, C. C., Giavalisco, M.,
  Dickinson, M., \& Adelberger, K. L. 1996, \aj, 112, 352
\bibitem[Steidel et~al.(1998)]{steidel1998} Steidel, C. C., Adelberger, K. L.,
  Dickinson, M., Giavalisco, M., Pettini, M., \& Kellogg, M.  1998, \apj,
  492, 428
\bibitem[Steidel et~al.(1994)]{steidel1994} Steidel, C. C., Dickinson, M.,
  \& Persson, S. E. 1994, \apjl, 437, L75
\bibitem[Teplitz et~al.(1998)]{teplitz1998} Teplitz, H. I., et al. 1998,
  BAAS, 193, 7507
\bibitem[Tytler \& Fan(1992)]{tytler1992} Tytler, D.  \& Fan, X. -M.
  1992, \apjs, 79, 1 
\bibitem[Turnshek(1984)]{turnshek1984} Turnshek, D. A. 1984, \apj, 280, 51
\bibitem[Vanden Berk et~al.(1999)]{vandenberk1999} Vanden Berk, D. E.,
  et al. 1999, \apjs, 122, 355
\bibitem[V{\' e}ron-Cetty \& V{\'e}ron(1996)]{veron1996} V{\' e}ron-Cetty,
  M. P., \& V{\' e}ron, P. 1996, ESO Sci.  Rep., 17, 1
\bibitem[Warren et~al.(1991)]{warren1991} Warren, S. J., Hewett, P. C.,
  Irwin, M. J., \& Osmer, P. S. 1991, \apjs, 76, 1
\bibitem[Vogt et~al.(1994)]{vogt1994} Vogt, S. S., et al. 1994, Proc. SPIE,
  2198, 362
\bibitem[Williams et~al.(1996)]{williams1996} Williams, R. E., et al. 1996,
  \aj, 112, 1335
\bibitem[Williger et~al.(1996)]{williger1996} Williger, G. M., Hazard, C.,
  Baldwin, J. A., \& McMahon, R. G. 1996, \apjs, 104, 145
\bibitem[Zhan et~al.(1989)]{zhan1989} Zhan, Y., Koo, D. C., \& Kron,
  R. G. 1989, \pasp, 101, 631
\end{thebibliography}
